\documentclass[manuscript,11pt]{aastex}
\usepackage{epsfig}

\def\fexviii{Fe\,{\sc xviii}}
\def\fexxiv{Fe\,{\sc xxiv}}
\def\fexxv{Fe\,{\sc xxv}}
\def\fexxvi{Fe\,{\sc xxvi}}

\def\civ{C\,{\sc iv}}

\def\mgii{Mg\,{\sc ii}}

\def\civ{C\,{\sc iv}}

\def\mathv{\textbf{\em v}}
\def\mathB{\textbf{\em B}}

\def\cm{\ifmmode {\rm cm}^{-1} \else cm$^{-1}$ \fi}
\def\s{\ifmmode {\rm s}^{-1} \else s$^{-1}$ \fi}
\def\cc{\ifmmode {\rm cm}^{-3} \else cm$^{-3}$ \fi}
\def\cs{\ifmmode {\rm cm}^{-2} \else cm$^{-2}$ \fi}
\def\g{\ifmmode \gamma \else $\gamma$\fi}
\def\G{\ifmmode \Gamma \else $\Gamma$\fi}
\def\Gs{\ifmmode \Gamma~ \else $\Gamma~$\fi}

\def\gc{\ifmmode \gamma_{\rm c} \else $\gamma_{\rm c}$ \fi}
\def\sw{Schwarzschild~}
\def\gsim{\mathrel{\raise.5ex\hbox{$>$}\mkern-14mu
             \lower0.6ex\hbox{$\sim$}}}
\def\lsim{\mathrel{\raise.3ex\hbox{$<$}\mkern-14mu
             \lower0.6ex\hbox{$\sim$}}}
\def\simless{\mathbin{\lower 3pt\hbox
     {$\rlap{\raise 5pt\hbox{$\char'074$}}\mathchar"7218$}}}   %< or of order
\def\simmore{\mathbin{\lower 3pt\hbox
     {$\rlap{\raise 5pt\hbox{$\char'076$}}\mathchar"7218$}}}   %> or of order
\def\Msun{M_\odot}                                % solar masses
\def\4u{4U 1728--34}
\def\deg{^\circ}
\def\aa{\buildrel _{\circ} \over {\mathrm{A}}}
\newcommand{\Alfven}{Alfv$\acute{\rm e}$n~}

\def\pg{PG~1211+143}

\lefthead{et al.} \righthead{\4u}

\shorttitle{Spectral Analysis of UFOs with MHD-Wind}

\shortauthors{Fukumura et al.}

\begin{document}

%\title{Constraining the Ultra-Fast Outflows in \pg\ with Magnetically-Driven Winds}
\title{Magnetically-Driven Accretion-Disk Winds and Ultra-Fast Outflows in \pg}

\date{\today}

\author{\textsc{Keigo Fukumura}\altaffilmark{1,2},
\textsc{Francesco Tombesi}\altaffilmark{3,4}
\textsc{Demosthenes Kazanas}\altaffilmark{3},
\textsc{Chris Shrader}\altaffilmark{3,5},
\textsc{Ehud Behar}\altaffilmark{6},
\textsc{and}
\textsc{Ioannis Contopoulos}\altaffilmark{7} }

\altaffiltext{1}{Email: fukumukx@jmu.edu}
\altaffiltext{2}{James Madison University, Harrisonburg, VA 22807} \altaffiltext{3}{Astrophysics
Science Division, NASA/Goddard Space Flight Center, Greenbelt, MD
20771}
\altaffiltext{4}{Department of Astronomy and CRESST, University of Maryland, College Park, MD20742}
\altaffiltext{5}{Universities Space Research Association, 7178 Columbia Gateway Dr. Columbia, MD 21046} \altaffiltext{6}{Department of
Physics, Technion, Haifa 32000, Israel} \altaffiltext{7}{Research Center for Astronomy, Academy of
Athens, Athens 11527, Greece}

\begin{abstract}

\baselineskip=15pt

We present a study of X-ray ionization of magnetohydrodynamic (MHD) 
accretion-disk winds in an effort to constrain the physics
underlying the highly-ionized ultra-fast outflows (UFOs) inferred by X-ray
absorbers often detected in various sub-classes of Seyfert active
galactic nuclei (AGNs). Our primary focus is to show that
magnetically-driven outflows  are indeed  physically plausible
candidates for the observed outflows  accounting for
the AGN absorption properties of the present X-ray spectroscopic
observations. Employing a stratified MHD wind launched across the
entire AGN accretion disk, we calculate its X-ray ionization
and the ensuing X-ray absorption line spectra. Assuming an appropriate ionizing AGN  spectrum, we
apply our MHD winds to model the absorption features in an {\it
XMM-Newton}/EPIC spectrum of the narrow-line Seyfert, \pg. We
find, through identifying the detected features  with Fe K$\alpha$
transitions, that the absorber has a characteristic ionization
parameter of $\log (\xi_c \textmd{[erg~cm~s$^{-1}$]}) \simeq 5-6$
and a column density on the order of $N_H \simeq
10^{23}$ cm$^{-2}$, outflowing at a characteristic velocity of
$v_c/c \simeq 0.1-0.2$ (where $c$ is the speed of light). The
best-fit model favors its radial location at $r_c \simeq 200 R_o$
($R_o$ is the black hole innermost stable circular orbit), with an
inner wind truncation radius at $R_{\rm t} \simeq 30 R_o$.
The overall K-shell feature in the data is suggested to be dominated
by \fexxv\ with very little contribution from \fexxvi\ and
weakly-ionized iron, which is  in a good agreement with a series of
earlier analysis of the UFOs in various AGNs
including \pg.

%This analysis in the context of our MHD-driven disk-winds can be
%generalized to other WAs %and UFOs in AGNs towards a systematic understanding
%of global outflows.

\end{abstract}


\baselineskip=15pt

\section{Introduction}

Blueshifted absorption lines are among the most common spectral
features seen in the spectra of accreting compact objects across a
large dynamic range in black hole (BH) mass from the upermassive BHs
of active galactic nuclei (AGNs) to stellar-mass black holes of
galactic binary systems. In the former case,
%
%Winds emanating from objects thought to be powered by accretion onto
%a compact object (black hole, neutron star or white dwarf) are
%common occurrence, manifest as blue shifted absorption features in
%their spectra.
%
approximately 50\% of Seyferts and quasars (QSOs) exhibit absorption
signatures in the UV band \citep[e.g.][]{Crenshaw99} with a similar
fraction ($\sim 50\%$) of Seyfert 1's showing blueshifted absorption
features in their X-ray spectra \citep[][]{Reynolds97,George98}
indicative of an underlying physical link between these two outflow
components. A small fraction ($\sim 10\%$) of the radio-quiet QSOs
further shows substantially blueshifted UV resonance lines, referred
to as broad-absorption-lines (BALs); these are mainly \civ/\mgii\
(high/low-ionization) at velocities of $v/c \sim 0.04-0.1$ where $c$
is speed of light \citep[e.g.][]{CKG03}.

X-ray spectroscopy plays a fundamental role in the study of AGN
absorber properties, because, in distinction to the UV transitions,
the X-ray ones span a much wider range in ionization
% instruments,
%including state-of-the-art gratings on board {\it Chandra} and {\it
%XMM-Newton} observatories, are a particularly powerful tools for
%probing the physical conditions of absorbing plasma outflows since
%they can cover a wide range of ionization
parameter $\xi$ (i.e. the ratio of photon to electron fluxes); thus,
within the span of 1.5 decades in photon energy ($\sim 0.3 - 10$
keV) one can sample atomic transitions that cover 5 decades in $\xi$
(e.g. from neutral to \fexxvi) and presumably a large range in
physical length scales. Typically, the so-called X-ray warm
absorbers (WAs) have characteristic local columns of $N_H \lesssim
10^{22}$ cm$^{-2}$ and ionization parameter in the range $-1
\lesssim \log \xi \lesssim 4$ at LoS velocities of $v/c \lesssim
0.01$ \citep[e.g.][]{ReynoldsFabian95} presumably originating from
sub-pc to pc scales. A rich spectral diversity in soft X-ray regime
($\lesssim 2-3$ keV) with a large number of ionic transitions
affords the statistical studies of their X-ray absorption line
properties \citep[e.g.][]{Behar03}. Among them, the absorption
measure distribution (AMD) can be used as a global measure of the
density of the  radiation-absorbing gas along a line-of-sight (LoS).
AMD is the differential hydrogen-equivalent column $N_H$ per decade
of $\xi$, i.e. $d N_H/d \log \xi$, and it is computed from the
measured columns of a variety of ionic species of several elements
spanning a large range in $\xi$
\citep[e.g.][]{Steenbrugge05,HBK07,B09,Detmers11}. The AMD
determination in a number of radio-quiet Seyferts seems to indicate,
to zeroth order, a similar global column distribution (i.e. roughly
a constant AMD), implying a wind density $n(r)$ that is similar in
all of them and decreases like $n(r) \propto r^{-1}$ with radius $r$
\citep[e.g.][]{Detmers11,Holczer12}.

Furthermore, in recent years, exhaustive X-ray studies of Fe K-shell
transitions in AGNs by {\it XMM-Newton} and {\it Suzaku}
%
%and the grating spectrometers on board
%
have revealed the presence of another outflowing component in the
Seyfert spectra, typically identified as highly-ionized high-Z ions
such as \fexxv/\fexxvi\, with H-equivalent columns of $N_H \gtrsim
10^{23}$ cm$^{-2}$, high ionization parameter ($\log \xi \gtrsim 4$)
at near-relativistic outflow speeds of $v/c \gtrsim 0.03$, named for
this reason ultra-fast outflows (UFOs). The detected UFOs appear to
be ubiquitous across both radio-quiet Seyferts like \pg\ \citep[see
also][]{Reeves09,Pounds03,PoundsPage06,T10a,T11a,T12a,
Gofford13,Gofford14} and radio-loud ones (e.g. 3C~111 and 3C~120,
3C~390.3, 3C~445) with a likely association of their
properties to the radio spectra \citep[][]{T10b,T11b}. The
higher X-ray content and increased wind ionization of the
latter suggests that strong X-ray photoionization apparently does
not inhibit the launch of such fast winds.
A detailed study of the properties of X-ray absorbers in a
sample of 23 AGN using high resolution X-ray spectroscopy was
conducted by \cite{Blustin05}, to conclude that most of the X-ray
absorbing matter is launched from large radii (the AGN molecular
torus) with kinetic luminosities that are only a small fraction of
the AGN budget. % in particular conducted a multi-parameter survey of
%X-ray absorbers from a sample of 23 AGNs to derive phenomenological
%and physical properties of their warm absorbers using the results of
%high-resolution X-ray spectroscopy.
%
In addition to these Seyferts and nearby QSOs, optically/UV-bright
broad-absorption-line (BAL) QSOs and their variants (e.g. non-BAL
and mini-BAL QSOs) apparently show similar X-ray UFOs but with even
higher velocities, up to $v/c \sim 0.7-0.8$ in extreme
cases\footnote[1]{While the winds of the typical high-velocity UV
transition (L$\alpha$, \civ\, etc.) BALs may be driven by line
radiation-pressure, it is a challenge for this scenario to
accelerate the highly-ionized, near-relativistic X-ray UFOs.} such
as APM~08279+5255
\citep[][]{Chartas02,Chartas03,Chartas07,Chartas09}.

\pg\ is  a bright quasar at redshift $z = 0.0809$
\citep[][]{Marziani96}  with X-ray luminosity $\sim 10^{44}$
erg~s$^{-1}$ in $2-10$ keV band for $H_0 = 75$ km~s$^{-1}$
Mpc$^{-1}$ and Galactic hydrogen-equivalent column density $N_H =
2.85 \times 10^{20}$ cm$^{-2}$ (Murphy et al. 1996). It is an
optically bright quasar with a prominent ``Big Blue Bump" that
results in a relatively steep optical/UV-to-X-ray flux
ratio\footnote[2]{The spectral index $\alpha_{\rm OX} \equiv 0.384
\log (f_{\rm 2 keV} / f_{\rm 2500})$ measures the X-ray-to-UV
relative brightness where $f_{\rm 2 keV}$ and $f_{2500}$ are,
respectively, 2 keV and 2500 $\aa$ flux densities (Tananbaum et al.
1979).}  ($\alpha_{\rm OX}= -1.45$).
%(based on the use of the NASA/IPAC Extragalactic Database or NED).
%It is unusual in the PG sample of bright quasars in having relatively
%narrow permitted optical emission lines (Boroson and Green 1992, Kaspi 2000).
%
%The first X-ray observation of \pg\ was made with {\it Einstein}
%showing a steep spectrum in the $\sim 0.2-2$ keV band (Bechtold et
%al. 1997, Elvis et al. 1991). In combination of {\it EXOSAT} and
%{\it GINGA} observations, Saxton et al. (1993) later resolved a
%strong soft X-ray excess above a nonthermal X-ray continuum
%component with photon index $\Gamma \sim 2.1$. The {\it ASCA}
%spectrum of \pg\ further confirmed a variable soft excess component
%and constrained the emission region be within $\sim 10^{15}$ cm
%(yaqoob+94). A subsequent braid-band  analysis has been made to
%further study a broad Fe K$\alpha$ emission  line and its equivalent
%width (Reeves+97) and UV to soft X-ray band (Janiuk+01).
%
Among other spectral features, the first {\it XMM-Newton}/EPIC/RGS
observation of \pg\  in 2001 revealed the first evidence of a
highly-ionized UFO with mass flux and kinetic energy comparable to
that of the accretion mass rate and bolometric luminosity,
respectively \citep[][]{Pounds03,PoundsReeves09}, although we  note
that others have reached different conclusions  depending on how to treat the baseline continua \citep[][]{KaspiBehar06,Gallo13,Zoghbi15}. Their analyses detected several strong absorption features
identified as blueshifted K$\alpha$ transitions of C, N, O, Ne, Mg,
S and Fe. The properties of the Fe feature in particular,
imply a wind column density of $N_H \sim 5 \times 10^{23}$ cm$^{-2}$
at velocity $v/c \sim 0.08$ and ionization parameter $\log \xi \sim
3.4$. A second observation of \pg\ with {\it XMM-Newton}/EPIC/RGS
in 2004 \citep{PoundsReeves07} and in 2007 \citep{PoundsReeves09}
have again detected similar UFOs, implying their persistent presence
despite its highly variable X-ray spectra.
A more detailed spectral analysis of such outflows has been recently
performed using  the {\it XMM-Newton} data
\citep[e.g.][]{T10a,T11a,Pounds14} to confirm their presence, in
agreement with the earlier results. Finally, a more recent
observation with {\it Suzaku}/XIS has also revealed the
same UFOs as well \citep[][]{Reeves08, Patrick12,
Gofford13}.

Despite the long-known UV/X-ray WAs combined with an increasing
number of statistically-significant detections of the X-ray UFOs,
the detailed geometrical structure of these ionized winds, including
formation and acceleration processes, are poorly constrained to
date. Yet, each of these issues is crucial to the comprehensive
picture of accretion-powered phenomena in accretion/outflow physics.
Plausible launching mechanisms for general outflows include
radiation-driven \citep[e.g.][in the context of UV BALs in luminous
QSOs]{PSK00,PK04,Nomura13}, thermally-driven \citep[e.g.][]{BMS83}
and magnetically-driven (e.g. \citealt{BP82}; \citealt{KK94}, \citealt{CL94},
hereafter CL94; \citealt{Ferreira97}; \citealt{FKCB10a}, hereafter
FKCB10a; \citealt{FKCB10b}, hereafter FKCB10b; \citealt{F14},
hereafter F14). There have also been   hybrid models
\citep[e.g.][]{Everett05,Ohsuga09,Ohsuga11,Proga03} that attempt to
explain an AGN phenomenology associated with inflow and
outflow\footnote[3]{The derived values of large $\xi, N_H$ and $v$
of certain X-ray UFOs presumably originating from smaller radii are
a serious challenge against line-driven and thermally-driven
scenarios.}.
With increasingly improved fully-numerical schemes, various
extensive simulations have been  made in the context of the
disk-wind scenario for (i) magnetically driven
\citep[e.g.][]{Pudritz06,Fendt06,PorthFendt10,Murphy10,Stute14,StepanovsFendt14}
and (ii) radiation driven
\citep[e.g.][]{PK04,Nomura13,Higginbottom14,Hagino15}.
Although the acceleration mechanism(s) of the observed winds remains
uncertain, the magnetic origin seems to be favored over the
radiation pressure one according to the latest time-dependent
hydrodynamic simulations coupled with multi-dimension Monte Carlo
calculations for radiative transfer (e.g., \citealt{Higginbottom14},
but also see \citealt{Hagino15}) and from UV/X-ray observations
\citep[e.g.][]{Kraemer05,Everett05,CK07}. This may also be the case
for Galactic binaries \citep[e.g.][]{Miller06,
Miller08,King12,King14}.
%
%For example, \citet{StepanovsFendt14}, among others,
%have performed MHD simulations to investigate launching processes of
%outflows from magnetically-diffusive accretion disk for different
%disk magnetization.
%
One should note that certain phenomenological outflow
models, with emphasis on individual spectral features, such as the
Fe K-shell transitions, are able to reproduce the properties of
certain prominent transitions such as their EW and their LoS
velocity \citep[e.g.][]{Sim05,Sim08,Sim10,Tatum12} without providing,
however, a global dynamic wind perspective.
%
%Motivated  by observational signatures, some AGN outflow models are
%able to partially reproduce physical properties of Fe K-shell
%emission and absorption signatures  such as equivalent width and LoS
%velocity under a phenomenological set of outflow assumptions
%\citep[e.g.][]{Sim05,Sim08,Sim10,Tatum12} with their focus primarily
%on individual spectral features (i.e. Fe K-shell transitions) rather
%than a global implication \citep[e.g.][]{KK94,K12}.

To the best of our knowledge, none of the existing wind models,
whether semi-analytic or numerical, have so far been able to deliver
a practical prescription for the observed X-ray absorption features;
i.e. local properties (i.e. column, ionization state, velocity) of
the WAs and UFOs together with a global picture of the outflow
physics (i.e. density/ionization structure from smaller scale to
larger scales and geometrical properties as a whole). From
methodological viewpoint, most models fit the properties of
specific features, i.e. column and velocity, implementing
\verb"xspec"/\verb"xstar" to obtain the ionization parameter
and the velocity of the plasma associated with specific transitions,
with little concern on how these fit within a global model of the
AGN outflows.
%
%
%on the other hand, many fitting models with
%\verb"xspec"/\verb"xstar" are implemented for their efficient
%execution for statistical analysis, but they are essentially {\it
%fitting-driven models} rather than {\it model-driven ones} in that
%data is fitted with conveniently-defined {\it phenomenological
%models} to derive certain parameters.

%An early ASCA observation showed the soft excess to be variable, indicating a source region of ²1015cm (Yaqoob et al. 1994).
%The improved spectral resolution of ASCA further refined the broad-band X-ray description of PG1211+143 (Reeves et al. 1997), with evidence for a broad Fe K emission line of equivalent width (EW) $\sim 400-750$ eV at $\sim 6.4$ keV. A more recent study of the overall (IR to X-ray) spectrum of \pg\ has been published by Janiuk et al. (2001), considering in particular the strong emission in the UV and soft X-ray bands and proposing its origin in a warm optically thick ÔskinÕ on the accretion disc. This work also included an analysis of an extended RXTE observation in 1997, suggesting a cold reflection factor $R = \Omega/2\pi$, where $\Omega$ is the solid angle subtended by the reflecting matter, of order unity...

The spirit of our recent works (i.e. CL94; FKCB10a; FKCB10b;
\citealt{K12}; F14) has been exactly the opposite, in that we begin
with a global MHD wind model and use the X-ray spectroscopic
observations to determine the global properties of these winds. In
this paper we employ a similar philosophy in an attempt to model the
observed \fexxv\ UFO in \pg\ within the context of the well-defined
MHD-driven wind models referred to above. Our study allows us to
constrain explicitly some of the defining MHD wind parameters in the
spirit of a {\it model-driven approach}. Our deeper goal is a better
understanding of the underlying physical structure of the observed
winds from a global standpoint. Within this framework, WA and UFO
features are generically identified as belonging to the same wind
structure that spans the entire domain of the AGN accretion disk. We
briefly describe the essence of the MHD-driven winds in \S 2 along
with our methodology for constructing a grid of simulated line
spectra for subsequent data analysis. In \S 3 we show our
preliminary results based on a 60-ks {\it XMM-Newton}/EPIC spectrum
of \pg\ deriving the best-fit values for the primary model
variables. We summarize and discuss the implications of the model in
\S 4.

\section{Ultra-fast Outflows in Stratified MHD Disk-Winds}

\subsection{The Magnetized Disk-Wind Structure}

Following FKCB10a and FKCB10b for the computational prescription of
magnetically-driven disk-wind models under steady-state,
axisymmetric conditions, we seek new insight into their structure
from the observational data. We apply our model assuming the
observed X-ray UFO signatures in AGNs (i.e. \fexxv/\fexxvi\
resonance transitions)\footnote[4]{The model, however, is not
restricted to Fe K-shell transitions and can be extended in general
to include other ionic features detected in AGNs and black hole
binaries.} are produced by X-ray photoionization of MHD winds
launched off of an accretion disk. The detailed characteristics of
the model discussed in FKCB10a and FKCB10b  will be briefly
described here. Geometric and physical properties of the wind in the
model are primarily governed by two conserved quantities along a
wind streamline, namely the particle-to-magnetic flux ratio $F_o$
and angular momentum $H_o$. The former one, $F_o$, determines
predominantly the wind kinematics and the latter, $H_o$,  generally
dictates the global wind structure in the poloidal plane. The
fundamental quantity of axisymmetric MHD is the magnetic stream
function  $\Psi(r,\theta)$, assumed to have a self-similar form
$\Psi(r,\theta) \equiv (R/R_o)^q \tilde{\Psi}(\theta) \Psi_o$, with
$\Psi_o$ the poloidal magnetic flux through the fiducial innermost
disk radius at $R=R_o$.
$\tilde{\Psi}(\theta)$ is its angular dependence to be
solved for and $q$ is a free parameter that determines the
radial dependence of the poloidal current.
The scalings of the poloidal magnetic stream function carry over to
the rest of the wind properties of which we show only the magnetic
field, velocity and density
\begin{eqnarray}
\mathB(r,\theta) &\equiv& (R/R_o)^{q-2} \tilde{\mathB}(\theta)B_o \ ,
\label{eq:eos} \\
\mathv(r,\theta) &\equiv& (R/R_o)^{-1/2} \tilde{\mathv}(\theta)v_o \ ,
 \label{eq:eos} \\
%     p(r,\theta) &\equiv& (r/r_o)^{2q-4} {\cal P}(\theta)B_o^2 \ ,
% \label{eq:pres2}  \\
n(r,\theta) &\equiv& (R/R_o)^{2q-3} \tilde{n}(\theta)B_o^2
v_o^{-2} m_p^{-1}\ ,
\end{eqnarray}
with the momentum-balance equation
\begin{eqnarray}
\rho (\bf{v} \cdot \nabla) \bf{v} &=& -\nabla p - \rho \nabla\Phi_g
+ \frac{1}{c} (\bf{J} \times \bf{B}) \ ,  \label{eq:momentum-eqn}
\end{eqnarray}
%
%.
where $m_p$ is the proton mass and $\rho$ is plasma mass density.

The dimensionless angular functions denoted by {\it tilde} must be
obtained from the conservation equations and the solution of the
Grad-Shafranov equation (the force balance equation in the
$\theta-$direction) with initial values on the disk (denoted by the
subscript ``o") at ($R=R_o, \theta = 90\deg$).
%
%It is more instructive to express the density normalization
%$(B_o/v_o)^2 m_p^{-1}$ in terms of the accretion or wind outflow
%rate $\dot m$, normalized to its Eddington value, as discussed in
%FKCB; then
%
The density normalization at ($R_o,90\deg$), setting
$\tilde{n}(90^{\circ})=1$, is given in terms of the dimensionless
mass-accretion rate $\dot{m}_a$ (normalized to the Eddington
accretion rate $\dot M_E = L_E/c^2$, see FKCB10a) by
\begin{equation}
n_o \equiv \frac{\tau(\dot{m}_a) f_w}{\sigma_T R_S} \ ,
\label{eq:no1}
\end{equation}
where $\sigma_T$ is the Thomson cross-section, $f_w$ is the ratio of
the outflow rate in the wind to $\dot{m}_a$ and $R_o$ is assumed to
be on the order of the \sw radius $R_S$. The Thomson depth
$\tau(\dot{m}_a)$ of the plasma at the innermost disk radius is
further scaled by the dimensionless mass-accretion rate $\dot{m}_a$
with normalization $\tau_o$ as $\tau(\dot{m}_a) \equiv \dot{m}_a
\tau_o$, which leads to
\begin{equation}
n_o \equiv \frac{f_w \dot{m}_{a,o}}{\sigma_T R_S} = 5
\left(\frac{f_w \dot{m}_{a,o}}{M_8} \right) \times 10^{11} ~ \textmd{cm$^{-3}$} \ .
\label{eq:no2}
\end{equation}
%
%It is important to note
%that because the mass flux in these winds depends in general on the
%radius, $\dot m$ always refers to the mass flux at the innermost flow
%radius at $r \simeq r_s$ where $r_s$ is the \sw radius.
%
where we have introduced an effective mass-accretion rate
$\dot{m}_{a,o} \equiv \dot{m}_a \tau_o$ as it is difficult to
decouple one from the other from observations. In this paper we
consider one of the fiducial wind solutions, model (A), from
\citet[][]{F14}, as a baseline wind model, by choosing $q=0.93$
(i.e. $n \propto r^{-1.14}$), $f_w=1$ and $\tau_o=10$ representing
an optically-thick disk of $\dot{m}_{a,o} = 10$ at $R=R_S$ (low case
$r$ denotes the radial distance in 3-space, while $R$ along the disk
surface) .
%The poloidal wind property is shown in Figure~\ref{fig:f1}a where colors indicate the normalized plasma density $\log n(r,\theta)$ superimposed with density contours (solid), poloidal velocity $v_p$ (arrows), velocity contours (dashed), and poloidal magnetic fields \mathBp\ ({\it thick solid} curves).
Here, we only highlight the essence of the model; details can be
found elsewhere (CL94;  FKCB10a,b; \citealt{K12}; F14).
Formally the self-similar winds extend from $r=0$ to $r \rightarrow
\infty$, however, physical considerations restrict these to a finite
but broad range in $r$. So we choose the  dimensionless factors
$f_t, f_T$ to denote the inner and outer truncation radii of our
winds on the disk surface by % The wind truncation radius $R_{\rm T}$
%in our model is normalized to the innermost disk radius $R_o$ by its
%normalization $f_T$ as
%
\begin{eqnarray}
R_{t} \equiv f_t R_S \ , \ R_{T} \equiv f_T R_S \label{eq:fr}
\end{eqnarray}
where the value of $f_t$ is to be constrained by the X-ray data
while $f_T \gg 1$, \textbf{typically $\sim 10^6$}. Once launched,
the asymptotic wind speed in this solution is found to be $v_p/v_o
\sim 4$ at $r/R_S \lesssim 10^3$ (see F14 for details).

\begin{figure}[ht]% ------------------------------------- Figure~1
\begin{center}$
\begin{array}{cc}
\includegraphics[trim=0in 0in 0in
0in,keepaspectratio=false,width=3.0in,angle=-0,clip=false]{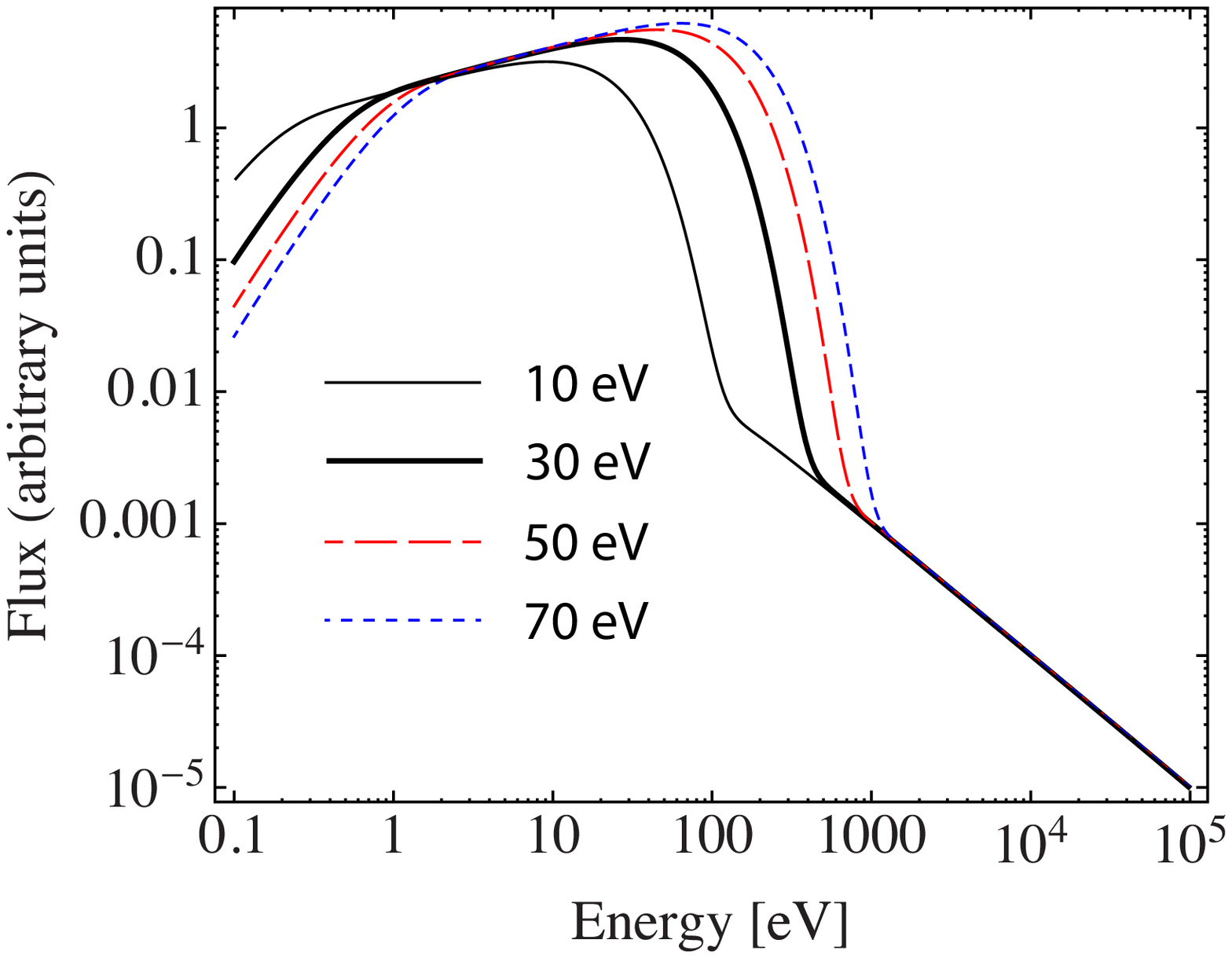} &
\includegraphics[trim=0in 0in 0in
0in,keepaspectratio=false,width=3.0in,angle=-0,clip=false]{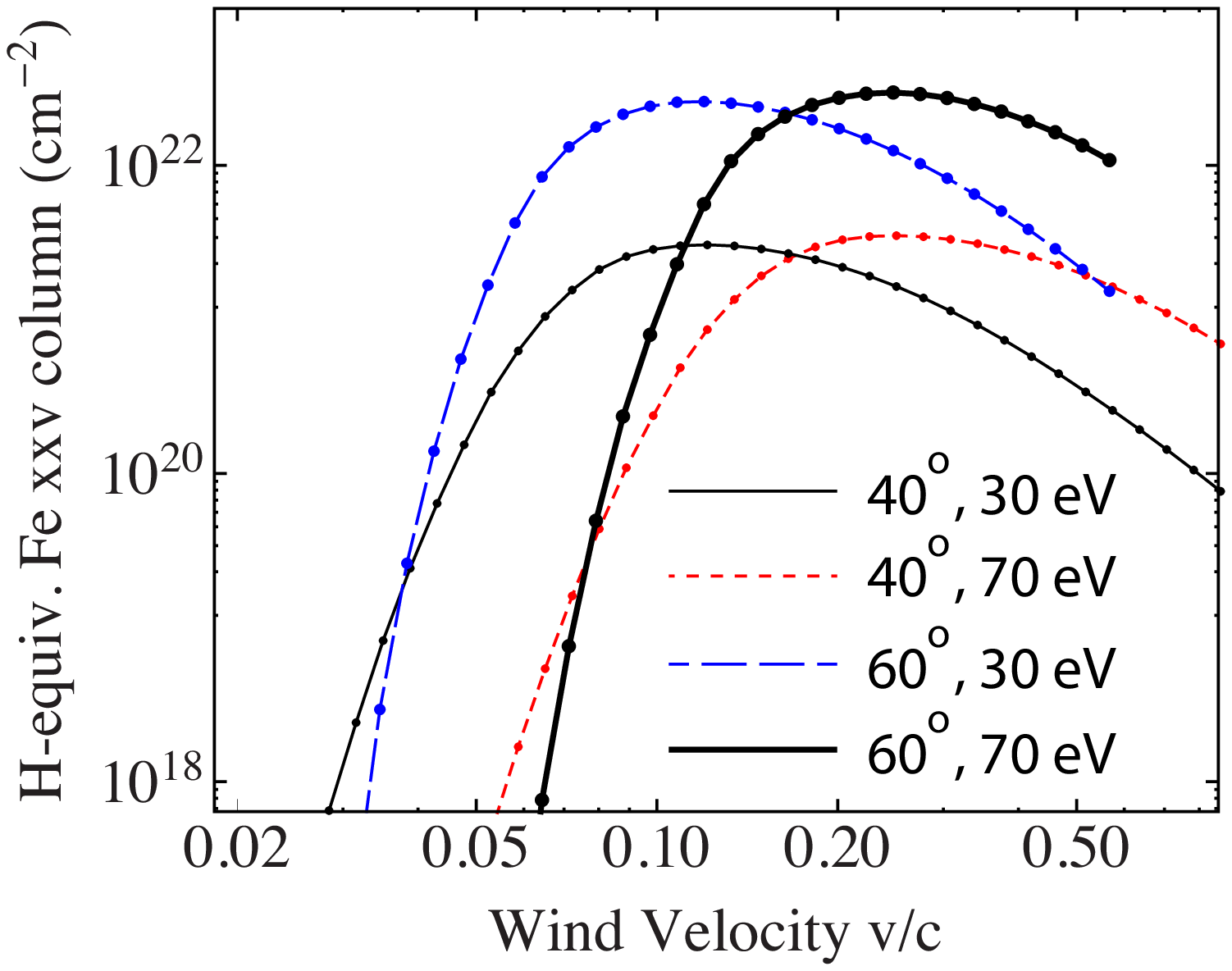}
\end{array}$
\end{center}
\caption{(a) Input ionizing SED for photoionization calculations for
the MCD of $kT_{\rm bbb}=10, 30, 50$ and $70$ eV with $\alpha_{\rm OX}=-1.5$
and $\Gamma=2$. (b) An example of calculated \fexxv\ (hydrogen-equivalent)
local column distribution ${\cal N}_H$ for different sets of $(\theta, kT_{\rm bbb})$
as labeled.} \label{fig:fig1}
\end{figure}

\subsection{Photoionization of Disk-Winds}

With the dimensionless, mass-invariant wind structure for a given
$\dot{m}_{a,o}$ and a viewing angle $\theta$, the only significant
difference in the wind ionization properties across objects of
different luminosity comes from the spectral energy distribution
(SED) of the accretion-powered luminosity $L \equiv \dot{m}_{a} L_E
\, \epsilon$ where $L_E = 1.25 \times 10^{46} M_8$ erg~s$^{-1}$ is
the Eddington luminosity with $M_8$ the black hole mass in units of
$10^8 \Msun$ and $\epsilon \simeq 0.1$ is the accretion efficiency.
While in FKCB10a we used a simple power-law spectrum of the form
$F_{\nu} \propto \nu^{-1}$ \citep[e.g.][]{Sim08,Sim10},
here we consider a multi-component SED consisting of a
multicolor-disk (MCD) with an innermost temperature of $kT_{\rm
bbb}$ and an X-ray power-law of photon index $\Gamma$ (with a
low-energy cut-off at 50 eV and a high-energy turnover at 200 keV)
normalized to the MCD by $\alpha_{\rm OX}$
\citep[e.g.][]{Everett05,Sim05}, a more appropriate SED for bright
Seyferts such as \pg. The ionizing luminosity (X-ray plus EUV) is
then $L_{\rm ion} \simeq 0.1 L \simeq 1.25 M_8 \times 10^{44}$
erg~s$^{-1}$ for a relatively high accretion rate of $\dot{m}_{a}=1$
as suggested in the earlier analyses \citep[e.g.][]{Pounds03}.

For a characteristic Seyfert SED, we set $\Gamma=2$ (see Fig.~5 in
\citealt{T11a} for a homogeneous sample of 42 radio-quiet AGNs and
\citealt{PoundsReeves09}) and $\alpha_{\rm OX}=-1.5$ (adopted from
NED and \citealt{Blustin05})  while leaving an inclination angle
$\theta$ and the disk temperature $kT_{\rm bbb}$ as free parameters
to be determined by \pg\ UFO observations
\citep[e.g.][]{Pounds03,T11a}.
It should be noted that, in agreement with FKCB10b,
\cite{Blustin05} find that only more negative values of $\alpha_{\rm
OX}$ allow higher velocity absorbers (i.e. $v_{\rm out} \gtrsim
10,000$ km~s$^{-1}$) based on their analysis of phenomenological and
physical properties of the detected warm absorbers using
high-resolution X-ray spectroscopy of a sample of Seyfert 1 type
AGNs.

Given the wind density normalization $n_o$ through $\dot m_{a,o}$,
photoionization balance is computed radially outward employing
\verb"xstar" \citep[][v2.2.1bn13]{KB01} by setting the SED of
Fig.~\ref{fig:fig1}a as the ionizing spectrum at the innermost
radius; the radiation transport in the wind is done by discretizing
the the radial wind coordinate using a large number of cells in
radial direction for a given angle $\theta$ (typically with $\Delta
r/r \sim 0.1$ allowing to treat each radial cell as a
plane, yielding 50-70 radial zones; see FKCB10a). We apply
\verb"xstar" in the first zone to compute the ionization equilibrium
of the plasma and also its opacity and emissivity. Then we use the
output of this zone as input for the next one and continue to the
outer edge of the wind
%Thus, a radiative transfer
%solution from one cell is provided as the subsequent initial
%condition moving to the next cell in an iterative manner from
%$r=r_{\rm in}$ all the way to $r=r_{\rm out}$
along a given LoS (i.e. a given $\theta$). We calculate the
absorption spectra with the Voigt function
\citep[e.g.][]{Mihalas78,Kotani00,Hanke09} defined as
\begin{eqnarray}
H(a,u) \equiv \frac{a}{\pi} \int_{-\infty}^\infty \frac{e^{-y^2}
dy}{(u-y)^2+a^2}   \ .\label{eq:voigt}
\end{eqnarray}
where we use $a \equiv \Gamma_E / (4 \pi \Delta \nu_D)$ with
$\Gamma_E$ being the Einstein coefficient and $\Delta \nu_D$ the
line Doppler broadening factor. The dimensionless frequency spread
about the transition frequency $\nu_o$ is given by $u \equiv (\nu -
\nu_0)/\Delta \nu_D$.  Note that, in order to compute the flux
in lines whose thermal width is narrower than the computational
frequency grid (especially in cases of multiple lines within a given
frequency spacing), the parameter \verb"vturb" (typically $\sim
1000$ km/s) of \verb"xstar" is employed to provide line width
$\Delta \nu_D$ consistent with the produced flux over the grid size.
 However, our wind model provides, instead, a well-defined velocity
shear $\Delta V$ with a corresponding radial velocity difference
$\Delta v_D$ between two adjacent radial cells; we employ this
velocity instead of \verb"vturb" to define an equivalent $\Delta
\nu_D = \nu_0 (v_D/c)\Delta v_D$, a value consistent with the
underlying wind kinematics (see FKCB10a for a detailed numerical
prescription).
%, in our view, a considerable improvement in the
%radiation transfer treatment of the specific problem.

Using the ionic column $N_{\rm ion}(r;\theta)$ over a radial cell of
width $\Delta r$ as a function of ionization parameter
$\xi(r;\theta)$ obtained with {\tt xstar} under ionization
and heating-cooling balance,  we can compute the wind opacity
$\tau_{\nu}(r,\theta)$ of any given photon energy
and at any given point with wind velocity $v(r;\theta)$
from the relation
\begin{eqnarray}
\tau_{\nu}(r,\theta) = \sigma_{\rm photo,\nu}(r,\theta)
N_{\rm ion}(r,\theta)  \ , \label{eq:tau}
\end{eqnarray}
where the line photoabsorption cross section $\sigma_{\rm
photo,\nu}$ at frequency $\nu$ is given by
\begin{eqnarray}
\sigma_{\rm photo,\nu} \equiv 0.001495 ~ \frac{f_{ij} H(a,u)}{\Delta
\nu_D} ~~\textmd{cm$^2$} \ , \label{eq:sigma}
\end{eqnarray}
and $f_{\rm ij}$ is the oscillator strength of the transition
between the i-th  and j-th levels of an ionic species.
Finally, we construct a two-dimensional grid of baseline spectra for
$\theta \in [30\deg, 70\deg]$  and $kT_{\rm bbb} \in  [10 {\rm eV},
70 {\rm eV}]$ for density normalization\footnote[5]{Note that the
wind density $n(r,\theta) \propto f(r)  g(\theta) n_o$ has both {\it
radial} $f(r)$ and {\it angular} $g(\theta)$ dependencies.
$g(\theta)$ (Fig.~2a of FKCB10a) decreases by factor of $\sim 10^4$ for
$0\deg \le \theta \le 90\deg$. } $n_o=5.1 \times
10^{11}$ cm$^{-3}$ ($\dot{m}_{a,o}=10$).
Here we introduce the quantity ${\cal N}_H$   defined as the
number density of \fexxv\ ions divided by the Fe abundance and
multiplied by the width of our local radial grid size $\Delta r$.
Some of the calculated  ${\cal N}_H$ (assuming solar abundances) for
four sets of $\theta$ and $kT_{\rm bbb}$ are shown as a function of
the wind velocity $v/c$ in Figure~\ref{fig:fig1}b.
Considering this figure, it is reminded that the
velocity decreases with increasing distance $r$ and decreasing
ionization parameter $\xi$ for a given LoS angle $\theta$.
The reader should note that  ${\cal N}_H$ does not depend
monotonically on velocity because, at small $r$ (and high $v$), a
good fraction of Fe is fully ionized, while at larger distances
(i.e. low velocities) the Fe ionization drops precipitously.
%
%Once a distance (velocity) is reached where the abundance of \fexxv\ is maximum
%while at larger distances with smaller velocities, the production of \fexxv\ is exponentially
%suppressed (matter much less ionized) and so does the H-equivalent
%column.
%
%We have plotted ${\cal N}_H$ for several scenarios in Figure~\ref{fig:fig1}b.
%
%\footnote[6]{Note that this column ${\cal N}_H$ actually denotes a
%hydrogen-equivalent \fexxv\ column over the LoS thickness $\Delta r$ within each computational
%``cell".}.
%
Finally, the {\em total} ${\cal N}_H$, $N_H($\fexxv), is found by
integrating ${\cal N}_H$ over $r$ along a given LoS.
As seen, for a given $\dot m_{\rm a,o}$, the normalization of the
LoS column depends primarily on the inclination $\theta$ while the
location of peak ${\cal N}_H$  (i.e. where ${\cal N}_H$ is maximum),
for a given $\alpha_{\rm OX}$ and $n_o$, is mainly determined by the
disk temperature $kT_{\rm bbb}$. Such a correlation is also
discussed in FKCB10b.

%------------------------------- Table~1
\begin{deluxetable}{l|c}
\tabletypesize{\small} \tablecaption{Model Grid of {\tt mhdwind}
Component } \tablewidth{0pt}
\tablehead{Primary Parameter & Range  }
\startdata
Viewing Angle $\theta$ (degrees) & $30\deg, 40\deg, 50\deg, 60\deg, 70\deg$ \\
BBB Disk Temperature $kT_{\rm bbb}$ (eV) & 10, 30, 50, 70 \\
%Density Profile $n(r) \propto r^{-p}$ & $p = 0.8, 1, 1.5$ \\
Disk Truncation Radius  $\log f_t \equiv \log (R_{\rm t}/R_o)$  & $0, 0.3, 0.6, 0.9, 1.2, 1.5, 1.8$   \\
\enddata
\vspace{0.05in}
Assume $M_8 = 1, \alpha_{\rm ox}= -1.5$, $\Gamma=2, n_o = 5 \times 10^{11}$ cm$^{-3}$  and $L_{\rm ion} =1.25 \times 10^{44}$ erg~s$^{-1}$.
\label{tab:tab1}
\end{deluxetable}

\begin{figure}[ht]% ------------------------------------- Figure~2
\begin{center}$
\begin{array}{cc}
\includegraphics[trim=0in 0in 0in
0in,keepaspectratio=false,width=3.4in,angle=-0,clip=false]{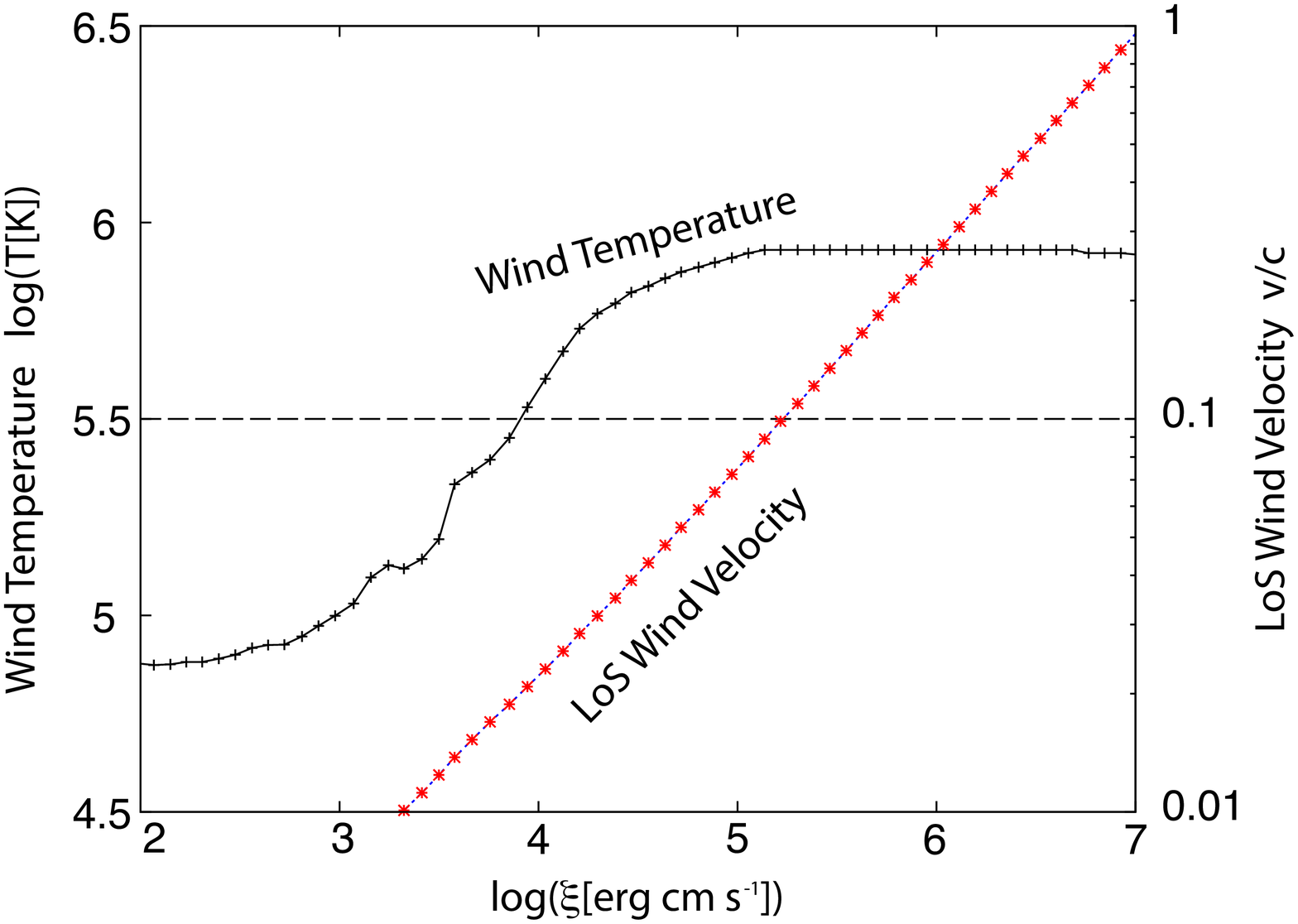}
\end{array}$
\end{center}
\caption{Radial profiles for photoionized wind temperature $\log
(T\textmd{[K]})$ (on the left ordinate) and velocity $v/c$ (on the
right ordinate) as a function of ionization parameter $\xi$ for
$\theta=50\deg$ and $kT_{\rm bbb}=30$ eV.  } \label{fig:fig2}
\end{figure}

\subsection{A Spectral Grid of {\tt mhdwind} Models}

\begin{figure}[ht]% ------------------------------------- Figure~3
\begin{center}$
\begin{array}{cc}
\includegraphics[trim=0in 0in 0in
0in,keepaspectratio=false,width=3.0in,angle=-0,clip=false]{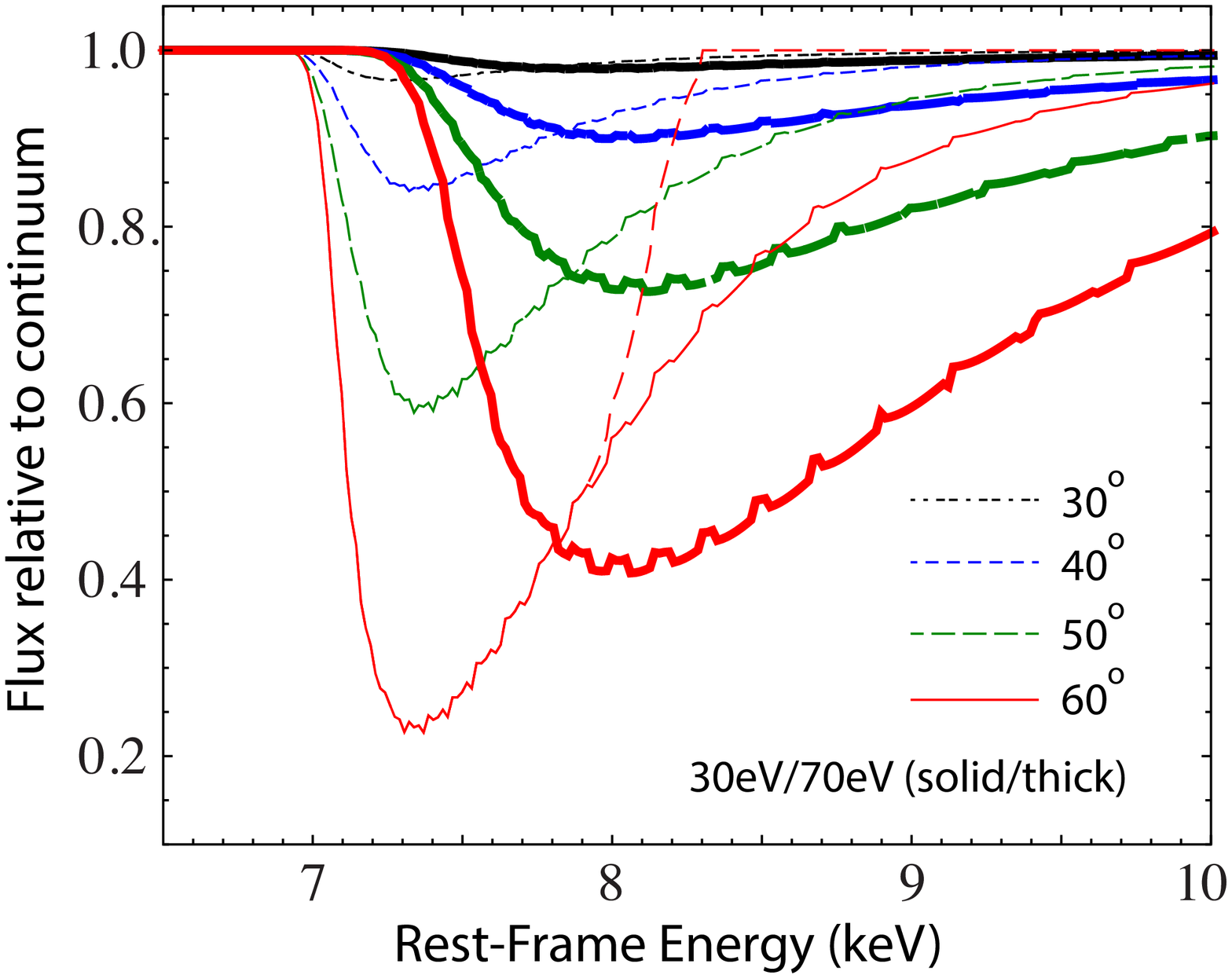}  &
\includegraphics[trim=0in 0in 0in
0in,keepaspectratio=false,width=3.1in,angle=-0,clip=false]{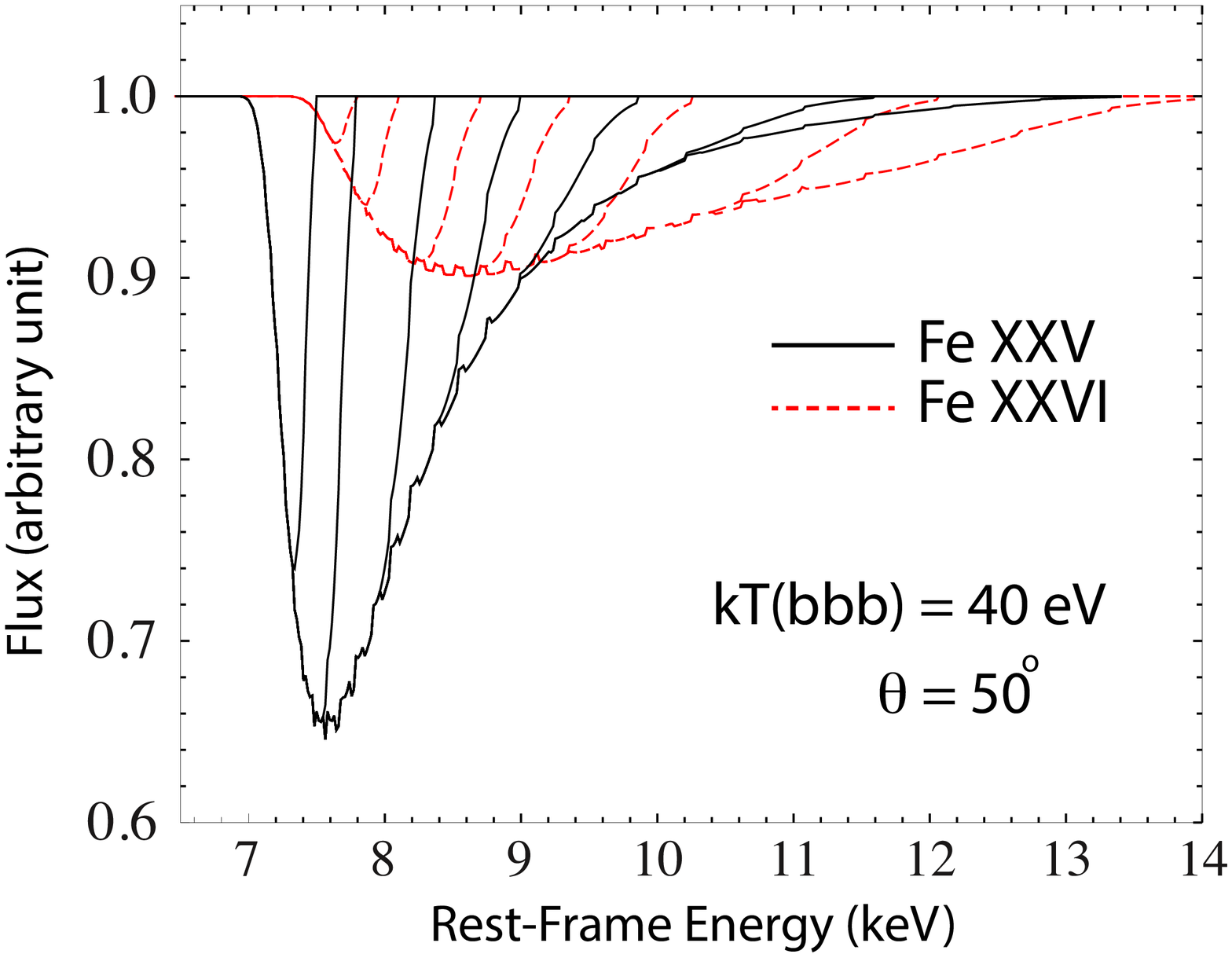}
\end{array}$
\end{center}
\caption{(a) Simulated \fexxv\ absorption features illustrating the diagnostic capability of our model  for $\theta=30\deg, 40\deg, 50\deg$ and $60\deg$ (from top to bottom) with thin curves denoting an input thermal disk spectrum with $kT_{bbb}=30$ eV and thick curves being $70$ eV. The dashed curve is given by $kT_{\rm bbb}=30$ eV but for $R_T/R_o = 10$. (b) A similar K-shell feature due to \fexxv\ (black) and \fexxvi\ (red) for various truncation radii $R_T$ as listed in Table~\ref{tab:tab1}. } \label{fig:fig3}
\end{figure}

For the model described in \S 2.1-2.2 and with the density
normalization fixed at $n_o=5 \times 10^{11}$ cm$^{-3}$ at
$\theta = 90^{\circ}$ (i.e. $\dot{m}_{a,o}=10$) we construct a grid
of {\tt mhdwind} models by varying  the three parameters $(\theta,
kT_{\rm bbb}, f_t)$. By solving the radiative transfer along
radial rays with the wind  photoionization as described
above we first obtain the column distribution for \fexxv\ and
\fexxvi\ in the ionized wind as a function of ionization parameter
$\xi(r,\theta)$ or equivalently \textbf{as} a function of
\textbf{the} wind velocity $v$ along a given LoS angle $\theta$ (see
FKCB10a,b) as shown in Figure~\ref{fig:fig1}b.
A different aspect of the wind is shown in Figure~\ref{fig:fig2}
where we demonstrate the wind velocity $v(\xi)$ and temperature
$T(\xi)$ under ionization balance for $\theta=50\degr$ and $kT_{\rm
bbb}=30$ eV. The wind velocity in this framework scales like $v
\propto r^{-1/2}$ as discussed in FKCB10a, FKCB10b and F14.
A sample grid of the simulated \fexxv\ absorption line profiles is
shown in Figure~\ref{fig:fig3}a for various combinations of
$(\theta, kT_{\rm bbb}, R_t)$ within the range considered here. As
seen, the dependence on each parameter can be probed by the spectral
shape in terms of the depth of trough and line shift. A comparison
between \fexxv\ and \fexxvi\ is shown in Figure~\ref{fig:fig3}b
where we explore the spectral variations for different truncation
radius $f_t \equiv R_t/R_o$ by setting $kT_{\rm bbb}=40$ eV and
$\theta=50\deg$.
One should note that the small fluctuations in the model line
profiles in Figure~\ref{fig:fig3} is not real and caused by the size
of the radial bin we choose in our model.

%\textbf{the following ({\em in em font}) has been discussed above
%and has to be removed}
%
%{\em Our wind model uses the velocity shear in the radial (i.e. LoS)
%direction, $\Delta v_D$, for the Doppler broadening $\Delta \nu_D$
%needed in Eqs. \ref{eq:voigt},\ref{eq:tau},\ref{eq:sigma}, thereby
%bypassing the need to introduce the parameter {\tt vturb} ($\simeq
%1000 - 5000$ km~s$^{-1}$) of \verb"xstar". This is equal to the
%velocity gradient between two adjacent computational cells and it is
%also the parameter used in computing the profiles of the absorption
%features. In our formalism, therefore, the spectral broadening is a
%natural element  rather than an arbitrary variable.
%
%
%Our wind model uses the velocity shear in the radial (i.e. LoS)
%direction $\Delta v_D$ for the Doppler-broadening  that is similar
%to velocity gradient between two adjacent computational cells in
%order to calculate the spectral line width. We stress here that the
%line width, therefore, is self-determined by $\Delta v_D$ by the
%wind kinematics which would otherwise be another free-parameter in
%typical modeling (i.e. turbulence or {\tt vturb}) arbitrarily
%assuming ${\tt vturb}=1,000-5,000$ km~s$^{-1}$. In our formalism the
%spectral broadening is thus a natural element  rather than an
%arbitrary variable.
%
%The calculated spectral features are consistent with the fact that
%the wind rapidly becomes tenuous at $\theta \lesssim 30\deg$ and
%also higher-temperature disk photons can more easily ionize Fe atoms
%to form \fexxv/\fexxvi\ at smaller radii (thus faster velocity or
%more blueshifted). }

%\begin{figure}[ht]% ------------------------------------- Figure~3
%\begin{center}$
%\begin{array}{cc}
%\includegraphics[trim=0in 0in 0in
%0in,keepaspectratio=false,width=3.0in,angle=-0,clip=false]{spec_grid2.eps}
%\end{array}$
%\end{center}
%\caption{An example of the simulated grid of \fexxv\ spectral lines for various combinations of $\theta=(30\deg, 60\deg)$, $\alpha_{\rm ox}=(-1.5, -1.7)$ and $f_r=(25, 50, 100, 200, 400, 800, 1600)$ for $\dot{m}_{\rm a,o}=1$ (i.e. $n_o=5 \times 10^{11}$ cm$^{-3}$). } \label{fig:fig3}
%\end{figure}

%\clearpage

\section{Preliminary Comparison with the \pg\ Data}

\subsection{{\it XMM-Newton}/EPIC Data}

%==[Francesco: can you check this paragraph???]

We use an {\it XMM-Newton} spectrum of \pg\ (obsID: 0112610101)
obtained with the EPIC-pn camera for an approximately $60$-ks
duration on 2001 June 15 \citep{Pounds03}, for which  a detailed
data reduction procedure and observed spectral and temporal features
of this object can be found elsewhere \citep[e,g,][reference
therein]{Pounds03,PoundsPage06,KaspiBehar06,T11a,Bachev09,PoundsReeves09,Gallo13,Pounds14}.
Earlier analyses of the UFOs, typically identified as either \fexxv\
and/or \fexxvi, seem to imply an estimate on the column density of
$N_H \sim 10^{23}$ cm$^{-2}$, velocity of $v/c \sim 0.1-0.15$ and
ionization parameter of $\log \xi \sim 3-5$, although an alternative
view may also be conceivable  claiming that the observed Fe K
absorption feature can be attributed to several consecutive low
charge states of Fe \citep[see, e.g.,][]{KaspiBehar06}.

\clearpage

%------------------------------- Table~2
\begin{deluxetable}{l|ccc}
\tabletypesize{\small} \tablecaption{Summary of our best-fit {\tt mhdwind} model parameters for \pg. } \tablewidth{0pt}  \tablehead{Parameter/Model & Model (A) &  & Model (B) \\
 & \fexxv/\fexxvi & & \fexxv/\fexxvi  }
\startdata
%$\dot{m}_{a,o}$ & 0.5 & 1  \\
%$n_{o}$[cm$^{-3}$]$/10^{11}$ & 2.5 & 5  \\ \hline
$\theta$ [degree] &  $40.0$ $^\diamondsuit$ &   & $49.8^{+3.27}_{-6.52}$  \\
$kT_{\rm bbb}$ [eV] &  $30.1^{+9.01}_{-2.56}$ &  & $38.1^{+4.55}_{-9.01}$  \\ \hline
%density & ? \\
%$r_{in}$ & ? \\
$E_{\rm Fe}$ [keV] & $6.54^{+0.097}_{-0.080}$ &  & $6.52^{+0.10}_{-0.073}$
\\ \hline
$\tau_{\rm max}$ & $0.095^{+0.014}_{-0.0105}/0.019^{+0.0023}_{-0.0031}$ &  & $0.235^{+0.073}_{-0.124}/0.052^{+0.018}_{-0.024}$  \\
$\log (r_c/R_S) ^\flat$ & $2.96^{+0.116}_{-0.161}/2.51^{+0.11}_{-0.16}$ &  & $2.37^{+0.48}_{-0.35}/1.82^{+0.54}_{-0.25}$  \\
$\log$ ($\xi_c$[erg~cm~s$^{-1}$]) $^\triangle$ & $5.21^{+0.149}_{-0.104}/5.62^{+0.147}_{-0.105}$ &  & $5.31^{+0.13}_{-0.15}/5.80^{+0.084}_{-0.17}$   \\
$v_c/c$ $^\triangle$  & $0.099^{+0.023}_{-0.008}/0.165^{+0.038}_{-0.013}$ &  & $0.115^{+0.016}_{-0.021}/0.208^{+0.018}_{-0.043}$  \\
%$\Delta N_{H,c}$(\fexxv) [cm$^{-2}$] /$10^{21}$ &  & & \\
$\cal{N}_H$ [cm$^{-2}$] /$10^{22}$ $^\sharp$ & $4.04^{+0.224}_{-0.178}/5.94^{+0}_{-0.182}$ &  & $12.1^{+5.30}_{-7.56}/16.7^{+5.72}_{-8.63}$  \\
$\log \left(R_{t}/R_S \right)$ & $0$ $^\clubsuit$ &  &
$1.48^{+0.065}_{-0.27}$
\\
$\chi^2/\nu$ (with \verb"mhdwind") & 200.84/129 &  & 198.54/128 \\
%$\chi^2/\nu$ (with $R_{\rm launch}/R_S=1$) &  & 233.56/128 \\
$\Delta \chi^2$ (from \verb"phabs*(po+zga)") & -34.1 &  & -36.4
\enddata
\vspace{0.0in}
\noindent
\begin{flushleft}
$^\diamondsuit$ The value is pegged. \\
$^\flat$ The characteristic LoS radius $r_c$ where wind \fexxv\  opacity $\tau_{\nu}$ (see eqn.~(9)) is maximum along a given LoS angle.  \\
$^\triangle$ The characteristic value (``c") is evaluated at the LoS position $r=r_c$.   \\
$\sharp$ LoS-integrated total \fexxv\ column density. \\
$^\clubsuit$ The value is fixed.
\end{flushleft}
\label{tab:tab2}
\end{deluxetable}

\subsection{Spectral Modeling for the Fe K$\alpha$ UFO}

Here we perform a spectral analysis of the UFOs previously detected
in the 60 ks {\it XXM-Newton}/EPIC spectrum of \pg\
\citep[e.g.][]{Pounds03,PoundsReeves09,T11a}. Focusing on the hard
X-ray absorption feature identified as Fe K-shell resonance
transition in the data, we implement our MHD-wind model,
\verb"mhdwind", into \verb"xspec" as a multiplicative table model as
discussed in \S 2. We follow the analysis procedure in  \citet{T11a}
where the $2-10$ keV band is modeled with an underlying continuum
power-law ({\tt po}). To fit the Fe K$\alpha$ absorber, however, we
replace the phenomenological {\tt xstar} component by our MHD-wind
model {\tt mhdwind}. The symbolic spectral form reads as ``{\tt
phabs*(po+zga)*mtable\{mhdwind\}}" where we have used the previously
estimated values of parameters,  Galactic absorption due to neutral
hydrogen column ({\tt phabs}) ${{N}}_H = 2.85 \times 10^{20}$
cm$^{-2}$ \citep[][]{Murphy96}, $\Gamma=2$ \citep[][]{T11a}, and
black hole mass, $M=10^8 \Msun$, based on the earlier estimatets
\citep[][]{Kaspi00,Bentz09}. Attributing the pronounced emission
line at $\sim 6.5$ keV (in the rest-frame)  to fluorescence
from the disk, our {\tt mhdwind} is constrained simultaneously with
a red-shifted gaussian component {\tt zga} of XSPEC in which the
line width is set to be $\sigma_{\rm Fe}=0.15$ keV whose exact value
has little influence on our end results.

\begin{figure}[h]% ------------------------------------- Figure~4
\begin{center}$
\begin{array}{cc}
\includegraphics[trim=0in 0in 0in
0in,keepaspectratio=false,width=3.0in,angle=-0,clip=false]{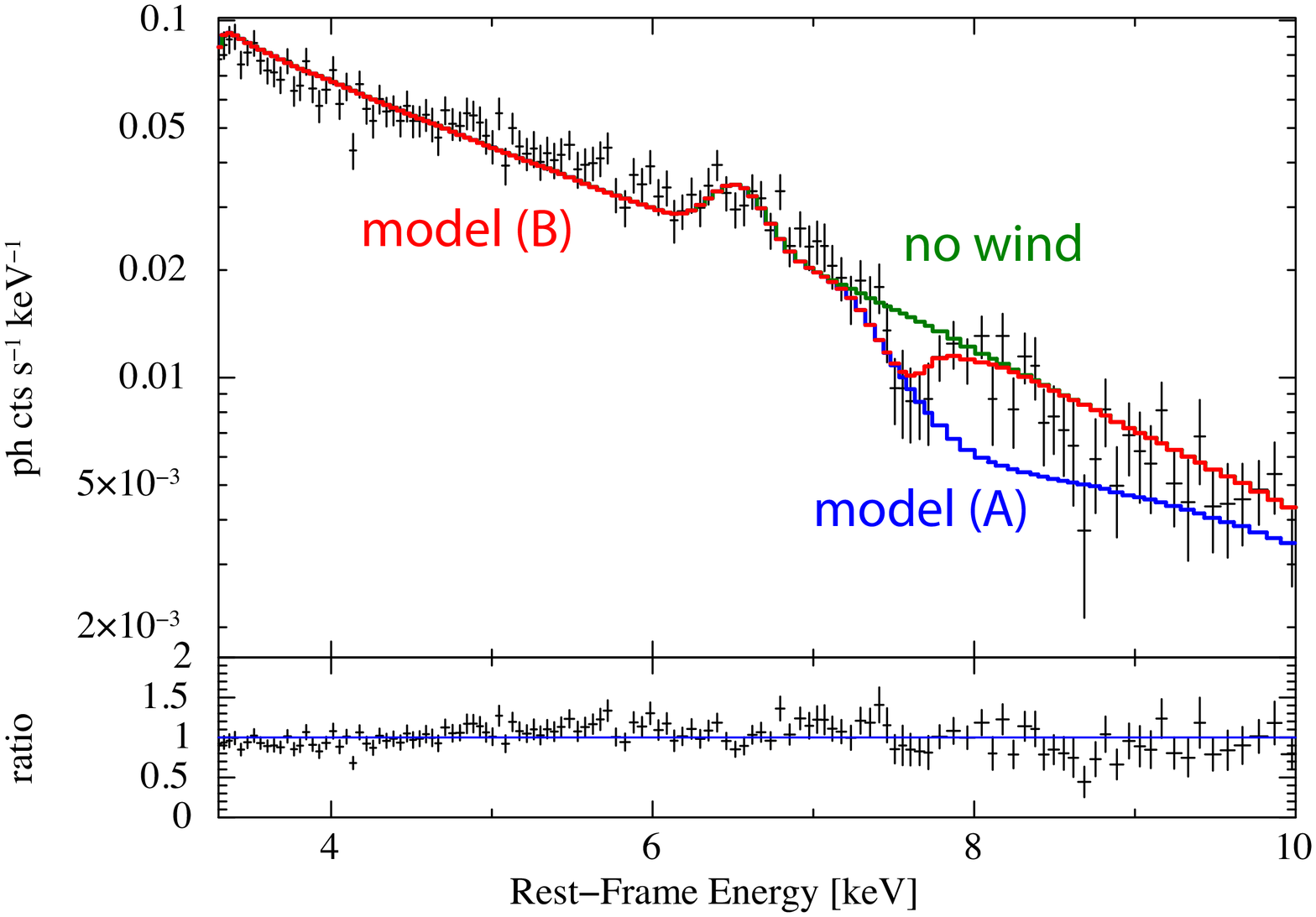} &
\includegraphics[trim=0in 0in 0in
0in,keepaspectratio=false,width=3.0in,angle=-0,clip=false]{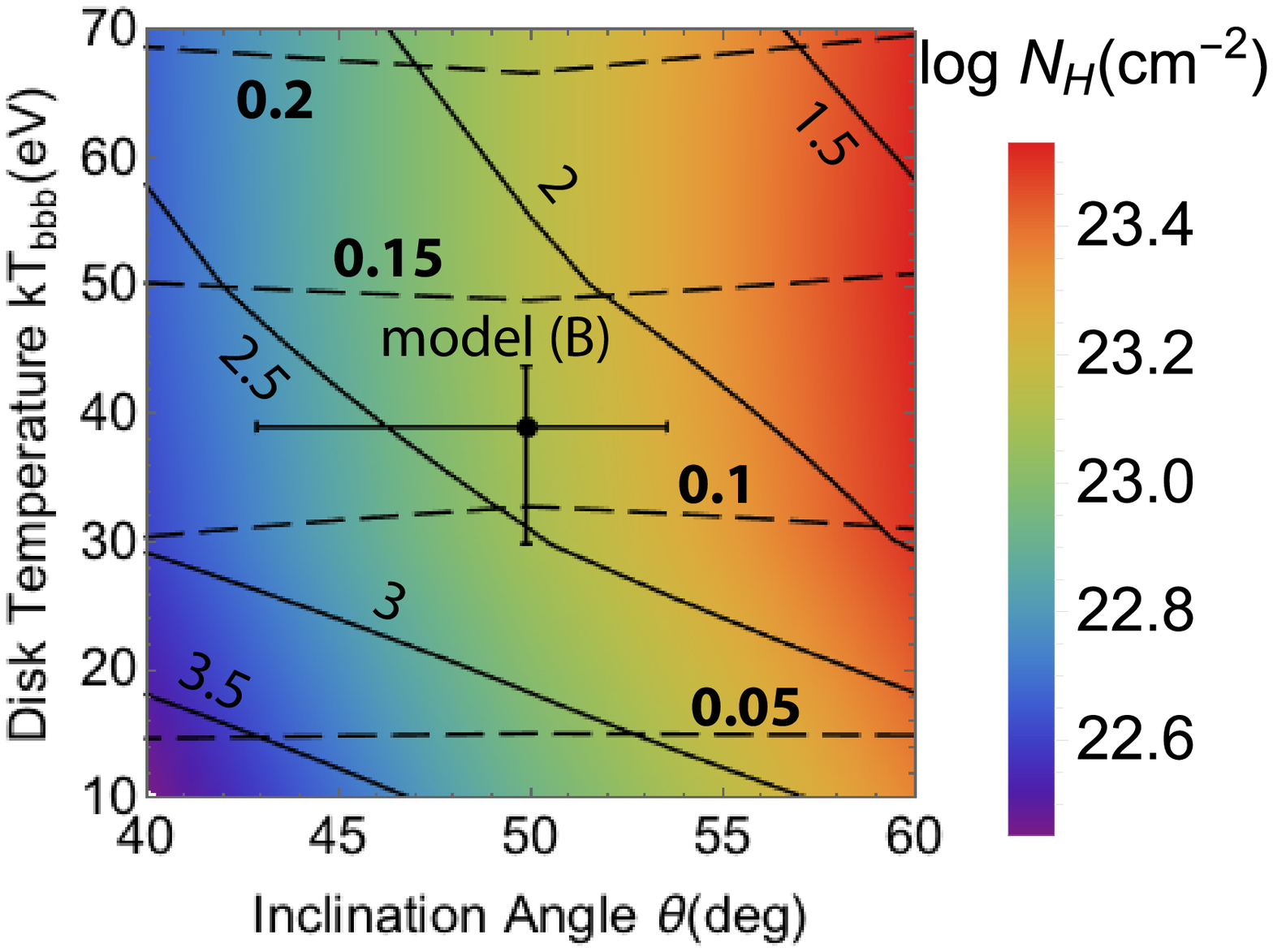}
\end{array}$
\end{center}
\caption{(a) 60 ks {\it XMM-Newton}/EPIC-pn spectrum of PG~1211+143
fitted with model (A) (blue) and (B) (red) in comparison with no
{\tt mhdwind} component (in green) varying three primary parameters
($\theta, kT_{\rm bbb}, f_t$) with $n_o =5 \times 10^{11}$ cm$^{-3}$
(i.e. $\dot{m}_{\rm a,o}=1$).  (b) Calculated contours for characteristic
radius $\log (r_c/R_S) = 1.5, 2, 2.5, 3, 3.5$ (solid; upper right to lower
left) and wind velocity $v_c/c=0.2, 0.15, 0.1, 0.05$ (dashed; top to
bottom) with the best-fit \fexxv\ column in model (B). Color indicates
the net Fe column density $\cal{N}_H$. See Table~\ref{tab:tab2} for details. }
\label{fig:fig4}
\end{figure}

%\subsubsection{With a Fixed Wind Launching Radius $r_{\rm launch}=r_o$}

We explore two cases by simultaneously considering both the  \fexxv\
and \fexxvi\ transitions: Model (A), where, $R_t$, the innermost
radial extent of the wind at $\theta = 90^{\circ}$, is
equal to $R_o \simeq R_S$
%coincides with  the fiducial non-truncated accretion disk
and model (B) where this restriction is relaxed.
Figure~\ref{fig:fig4}a shows the best-fit for each of
models (A) and (B) in comparison with the no {\tt mhdwind}
model. We set $n_o = 5 \times 10^{11}$ cm$^{-3}$ ($m_{\rm a,o}=10$)
while varying $\theta, kT_{\rm bbb}$ and $f_t$. The best-fit values
are listed in Table~2. We have used different values of $n_o$ in our
calculations but its effective role is to change slightly
the depth of the trough.
In the current wind model, both visual and  statistical inspection
favors model (B) by $\Delta \chi^2 = 2.3$ (table 2)  in which the
wind does not originate at the fiducial radius $R_o$ on the
disk surface, but at $R_t = f_t R_S$ with $f_t > 1$.

In model (A), where $f_t=1$ is assumed, we obtain our
best-fit for values $\theta=40\deg$ (pegged) and $kT_{\rm
bbb}=30$ eV with $\chi^2/\nu = 200.84/129$ with {\tt mhdwind} which
is a statistically significant addition to the continuum (with an
improvement of $\Delta \chi^2 = 34.1$ for two additional
parameters). %\textbf{({\em not too clear here})}
%Note that we only include \fexxv\ line in this preliminary work for simplicity.
In model (B) we relaxed the restriction on the wind
truncation radius $f_t = 1$ of model (A). Table~\ref{tab:tab3} shows
a list of various characteristic radii in this model.
Our analysis yields a best-fit model with $\theta=50\degr,
kT_{\rm bbb}=38$ eV and $f_t=10^{1.48}$,  as shown in
Figure~\ref{fig:fig4}a and Table~2, where we obtain  $\chi^2/\nu =
198.54/128$ which is more significant in comparison with
model (A). We note that the model spectrum now has a
sharp\textbf{er} edge on the bluer side of the feature as required
in data. The total column, ${\cal N}_H = 1.2 \times 10^{23}$
cm$^{-2}$ from model (B) is comparable to the previous estimate with
{\tt xstar} model although our wind is continuous rather than
discrete.
%\textbf{({\em the model B discussion was moved here from the
%previous page})}.

As a measure of assessing the \fexxv\ absorption wind
properties we first calculate a characteristic radius $R_c$  at
which the wind photoelectric absorption column for the \fexxv\
transition becomes maximum for a given LoS inclination angle
$\theta$. At this radius we compute the other physical quantities
listed in Table~2.
Notice that the {\em total} ${\cal N}_H$ (in units of
$10^{22}$ cm$^{-2}$) is defined as the local column density
integrated over the LoS distance.

%[[ Notice that the net (hydrogen-equivalent) Fe column $N_H$ (in
%units of $10^{22}$ cm$^{-2}$) is defined as the local column density
%integrated over the LoS distance]]\footnote[7]{Note that this column
%$N_H$ is the integrated binned column from Figure~\ref{fig:fig1}b.}.
%
%within which the contribution to the line flux is not negligible.
%
Along the LoS of the values of $\theta$ obtained by our fits
(see Table 2), the wind is both Thomson thin and also thin at the Fe
energies. Because, as argued earlier, the \fexxv/\fexxvi\ line
opacities are non-monotonic functions of the radial coordinate in
these directions, we define a radius $r_c$ along each of these LoS
at which the line(s) opacity (ies) is (are) maximized (given by the
entry $\tau_{\rm max}$ of Table 2). In fact, these coincide with the
maxima of ${\cal N}_H$ of Fig. 1b. Because of the smoothness and
continuity of ${\cal N}_H$ with $r$, $\xi$ or $v$, the absorption of
X-ray photons begins at $r \ll r_c$ and extends over more than one
decade in radius. One should hence bear in mind that a given
absorption feature in our models does not correspond to specific,
unique wind component.

%While the wind is globally optically-thin
%(i.e. $\tau(r) < 1$), it is found that the characteristic distance
%$R_c$ of the maximum optical depth $\tau_{\rm max} \equiv \tau(R_c)$
%roughly coincides with those of the maximum  \fexxv/\fexxvi\ columns
%along the LoS (see Fig.~\ref{fig:fig1}b). That also means that the
%actual X-ray absorption in the spectrum begins to take place at
%radius much smaller than $r=R_c$ (by orders of magnitude) since the
%column density  is a smooth and continuous distribution as a
%function of $\xi$ which is also a function of wind velocity $v$ (see
%FKCB10a). Therefore, the X-ray photons are progressively absorbed by
%different segments of the same wind along a LoS. With our choice of
%number density profile $n(r) \propto r^{-1}$, the column density is
%roughly constant per decade in distance.
%%
%A major feature of the MHD wind structure considered here is  the
%absence of a singly-characterized wind component. The wind has a
%well-defined gradient in its properties such as velocity, density
%and ionization parameter.

%------------------------------- Table~3
\begin{deluxetable}{l|c}
\tabletypesize{\small} \tablecaption{Various Characteristic Radii in
This Model} \tablewidth{0pt}
\tablehead{Symbol & Description  }
\startdata
$R_o$  & The fiducial radius of the wind models \\
$R_S$ & \sw\ radius \\
$R_t$ & Disk inner truncation radius  \\
$r_c$ & Radius where a given ion becomes most opaque  \\
$R_{\rm ISCO}$ & Innermost Stable Circular Orbit
\enddata
\noindent
\begin{flushleft}
All radii are defined in the cylindrical coordinates except for $r_c$ which is measured along a LoS direction. 
\end{flushleft}
\vspace{0.05in}
\label{tab:tab3}
\end{deluxetable}

%In model (A) the best-fit wind velocity is in the range of $0.04 \lesssim v/c \lesssim 0.06$. The required (integrated) column in the best-fit models is found to be $N_H< 10^{23}$ cm$^{-2}$. On the other hand, all the models require high values of ionization parameter $\log \xi \sim 4.5$ as commonly derived from the phenomenological {\tt xstar} modeling. All these models (A)-(C) are constrained based on the simplified situation where the launching disk radius $r_{\rm launch}$ coincides with the innermost disk radius $r_o$.
%We find that either model, (A)-(C), can explain data equally well at a very similar statistically significant level.

\begin{figure}[h]% ------------------------------------- Figure~5
\begin{center}$
\begin{array}{cc}
\includegraphics[trim=0in 0in 0in
0in,keepaspectratio=false,width=3.0in,angle=0,clip=false]{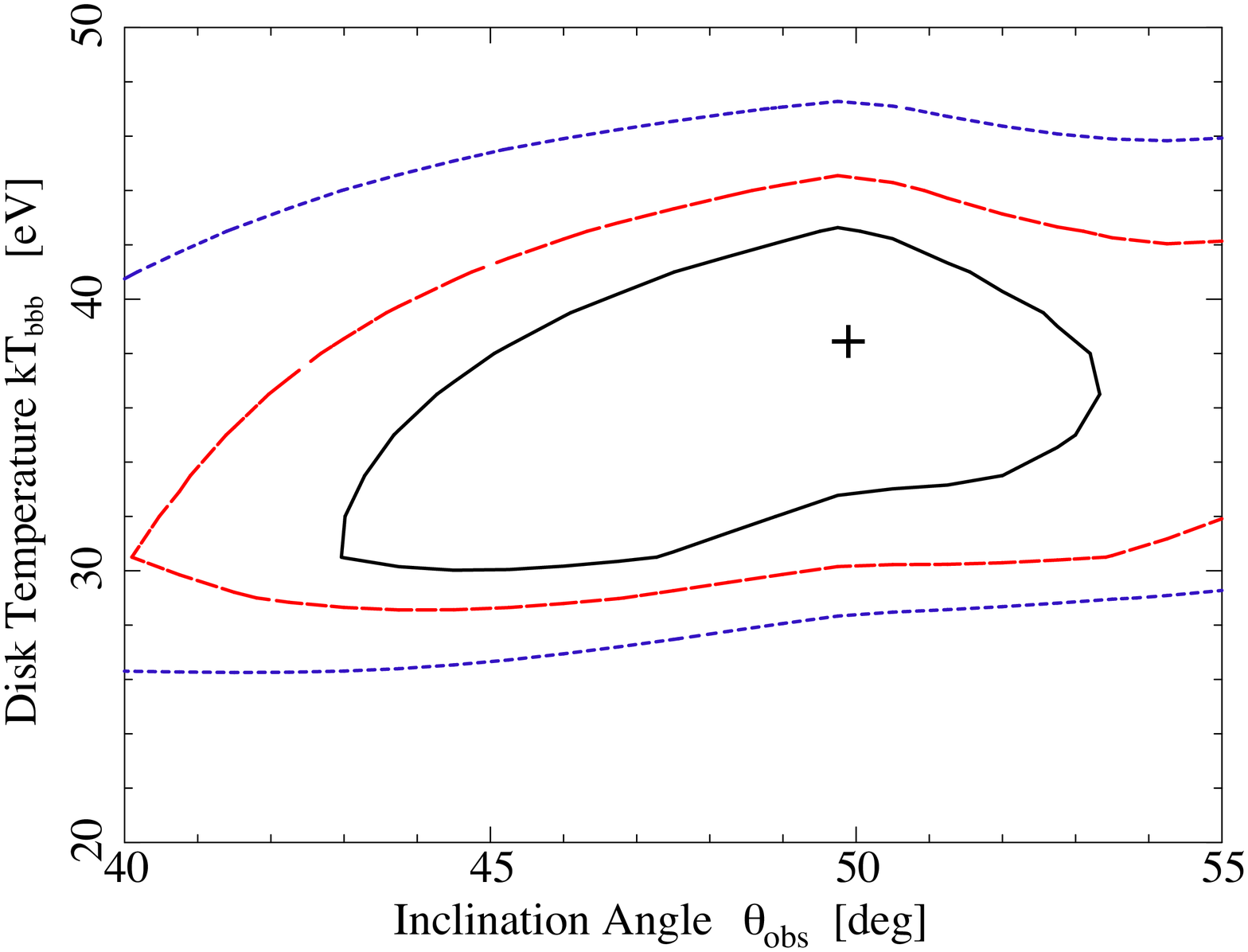} &
\includegraphics[trim=0in 0in 0in
0in,keepaspectratio=false,width=3.0in,angle=0,clip=false]{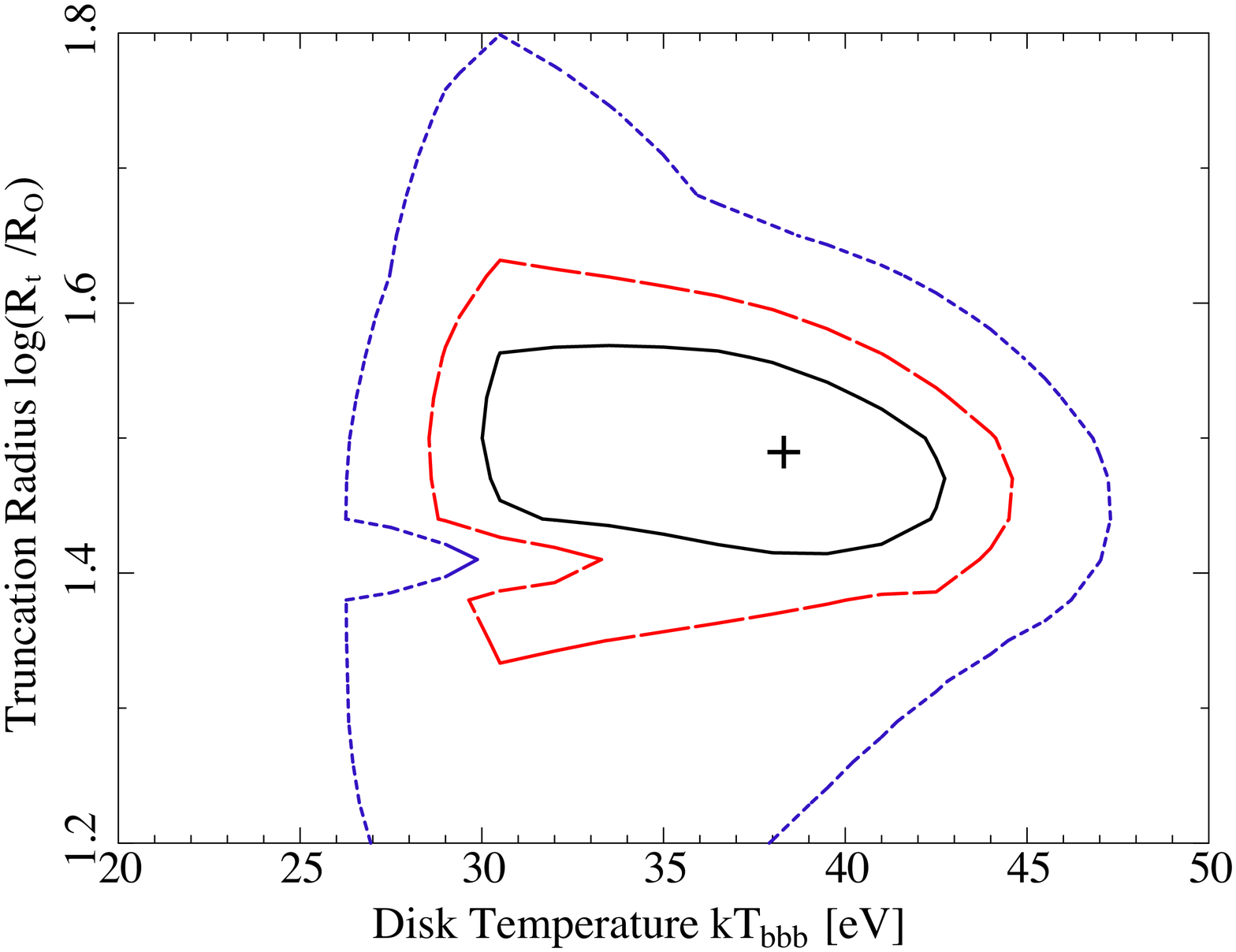}
\end{array}$
\end{center}
\caption{Confidence contour plots for (a)  temperature
$kT_{\rm bbb}$ and inclination $\theta$ and (b)  truncation radius $R_t$
and temperature $kT_{\rm bbb}$ at the confidence level of 68\%, 90\%,
and 99\%  also indicated by the best-fit model (B) (cross).
See Table~\ref{tab:tab2} for details. } \label{fig:fig5}
\end{figure}

%\textbf{I do not understand the next paragraph}
%
In order to examine in more detail the multi-parameter space spanned
by ($\theta, kT_{\rm bbb}$) we interpolate the wind variables such
as velocity $v_c$ and characteristic radius $r_c$ as shown in
Figure~\ref{fig:fig4}b where color shows total column $\log
(\cal{N}_H \textmd{[cm$^{-2}$]})$ for \fexxv\ with contours of
radius $\log (r_c/R_o)$ (solid), and  contours of $v_c/c$ (dashed).
The best-fit model (B) for \fexxv\ is indicated by a dark dot. It
should be reminded that the best-fit characteristic values (i.e.
$r_c, v_c, \xi_c$) are simply constrained at the most opaque radius
($\tau_{\rm max} \equiv \tau (r=r_c)$) of the absorber. The
neighboring plasma at $r \lesssim r_c$ (i.e. $v \gtrsim v_c$) and $r
\gtrsim r_c$ (i.e. $v \lesssim v_c$) contributes also progressively
to  the formation of the absorption feature thus there is no {\it
single} wind velocity nor column density in our model.  {\it This is
a characteristic feature of the continuous wind model which is
fundamentally different from a single-component absorber model often
employed in a phenomenological analysis. }
The corresponding confidence contours for model (B) are shown in
Figure~\ref{fig:fig5} where the primary variables $\theta, kT_{\rm
bbb}$ and $f_t$  are constrained.
%\textbf{it should be $f_t \equiv
%R_t/{\bf R_S}$ actually}

\begin{figure}[ht]% ------------------------------------- Figure~6
\begin{center}$
\begin{array}{cc}
\includegraphics[trim=0in 0in 0in
0in,keepaspectratio=false,width=3.0in,angle=-0,clip=false]{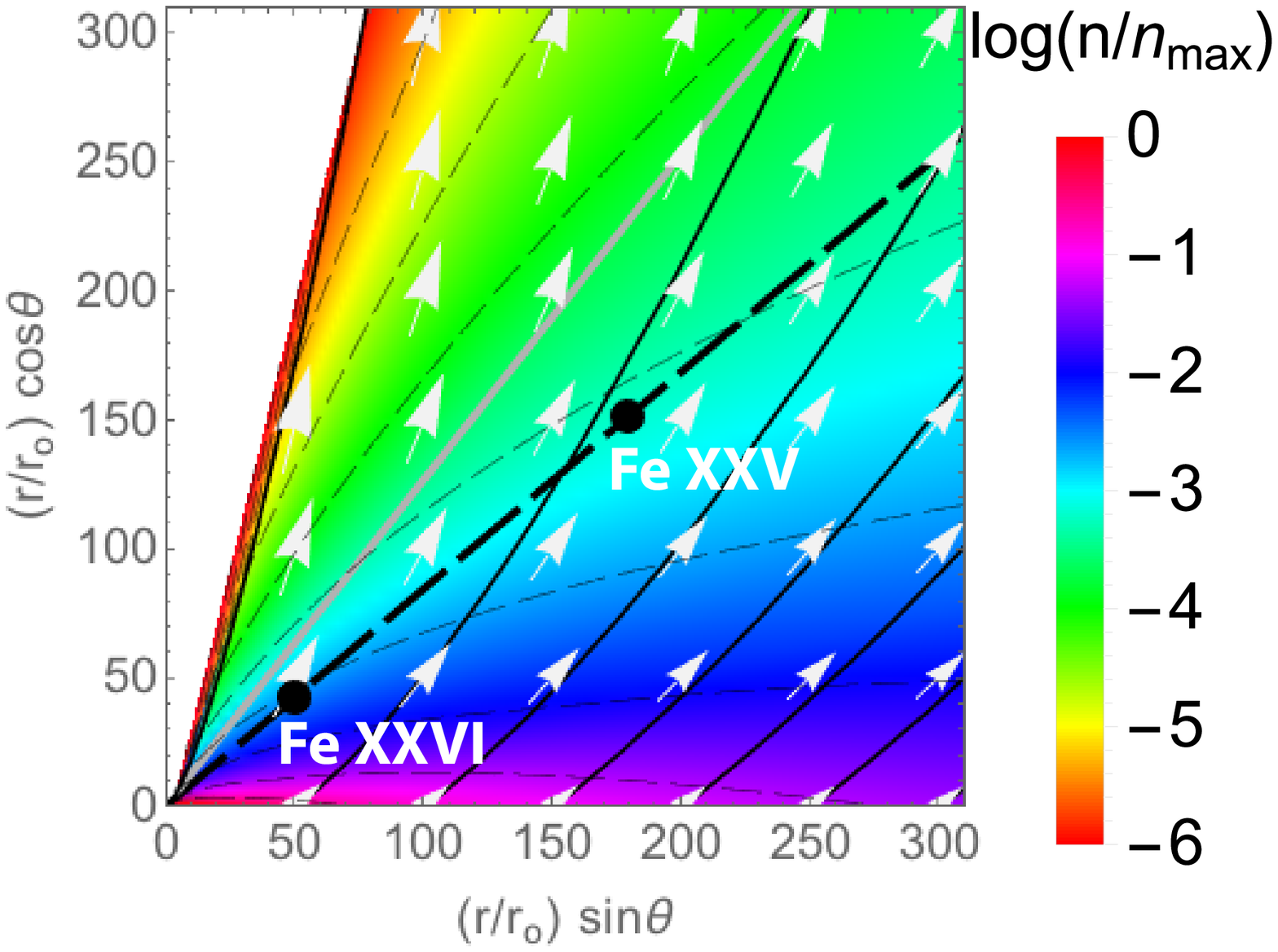}
&
\includegraphics[trim=0in 0in 0in
0in,keepaspectratio=false,width=2.9in,angle=-0,clip=false]{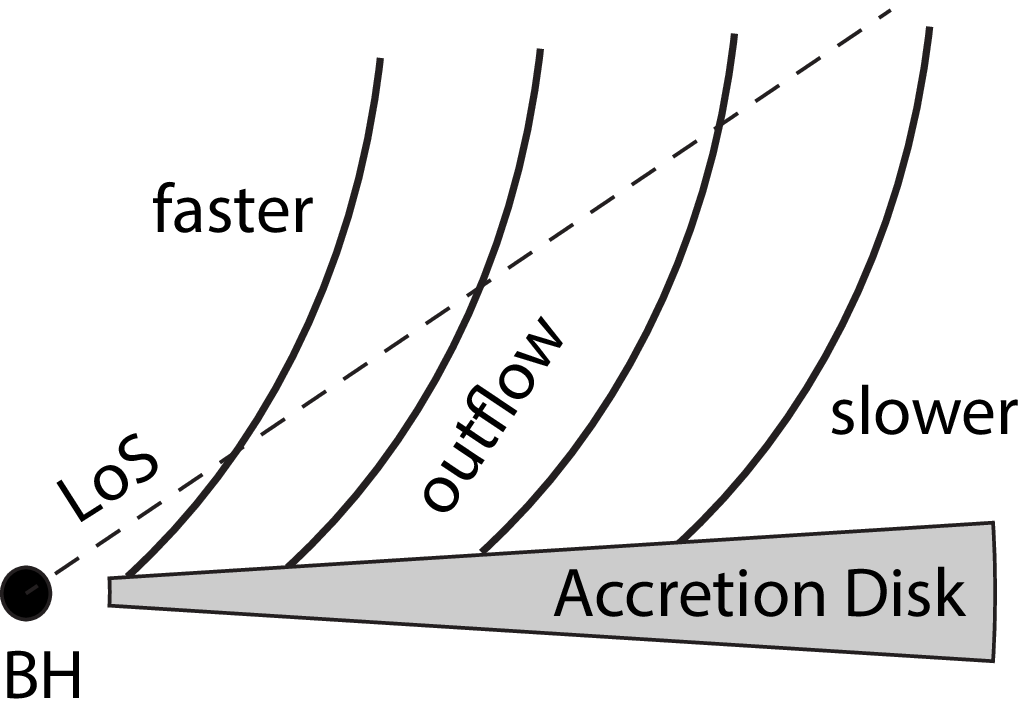}
\end{array}$
\end{center}
\caption{(a) Geometrical identification of the observed \fexxv/\fexxvi\
from {\it XMM-Newton}/EPIC spectrum of PG~1211+143 with {\tt mhdwind}
model (B). We show the global magnetic field lines (thick solid),
normalized wind density $\log (n(r,\theta)/n_{max})$ in color and
its contours (dashed), velocity field (white arrows) along with
the inferred location of \fexxv\ and \fexxvi. See Table~\ref{tab:tab2}
for details. (b) A schematic to illustrate the MHD-driven AGN winds. }
\label{fig:fig6}
\end{figure}

In the context of the current model, the best-fit model (B) is
spatially identified as illustrated in Figure~\ref{fig:fig6}a where
the calculated fiducial wind structure in the vicinity of the black
hole is superimposed showing the normalized number density
$n(r,\theta)$ in color, the velocity field (white arrows), the
magnetic field lines (solid thick), the contours for density
(dashed) and the \Alfven surface (white line). In this simplified
approach a geometrically-thin disk is situated in the equatorial
plane at $\theta=\pi/2$. As discussed earlier, the faster portion of
the modeled \fexxv/\fexxvi\ absorber (i.e. the bluer side of the
trough) and the slower one (i.e. the redder one) are respectively
located at $r < r_c$ and $r > r_c$ along each LoS and they all
progressively contribute to produce the observed absorption feature
(both in depth and width).

%All three models, (A)-(C), are compiled together in
%Figure~\ref{fig:fig6}b showing their spatial positions superimposed
%to the simulated MHD-winds (see model (A) from F14) in the poloidal
%view. We show (normalized) wind density distribution in color,
%density contours (dotted), a global magnetic field lines (solid),
%and poloidal velocity (arrows). The best-fit models are indicated by
%filled dots along different LoS (dashed). As discussed in
%Figure~\ref{fig:fig6}a the fastest portion of the modeled \fexxv\
%absorber (i.e. bluer side of the trough) is located at $r \ll r_c$
%along the constrained LoS angle.

In terms of the energy budget of the observed UFO,  by using the
outflow density profile with $n \propto 1/r$ (in the present model
actually $n \propto 1/r^{1.14}$) in this work
%
%and $\dot m =0.1$
%
for a supermassive black hole mass of $M=M_8$
\citep[][]{Kaspi00,Peterson04},  a mass-outflow rate associated with
the \fexxv\ line can be estimated as
%
%\footnote[3]{Note that the {\it intrinsic} mass-accretion rate
%$\dot{m}_{\rm net}$ is a sum of electromagnetically observable
%mass-accretion rate $\dot{m}_{\rm in}$ and outflow rate
%$\dot{m}_{\rm out}$ where the two quantities can be independent of
%each other.}
%
\begin{eqnarray}
\dot{M}_{\rm out}(\textmd{\fexxv})  &\equiv& 2 \pi b m_p  \int_0^{200}
n(r,\theta) v_z(r,\theta) r dr \ , \\
&\sim& 4 \pi b m_p n_o c R_S^2 v_{\rm \fexxv}  x_{\rm \fexxv}^{1/2} \ , \\
&\sim& 2.56 \Msun~\textmd{yr$^{-1}$} \left(\frac{b}{0.4}\right) \left(\frac{n_o}{5 \times 10^{11}} \right) \left(\frac{M}{10^8 \Msun}\right)  \left(\frac{v_{\rm \fexxv}}{0.1}\right)  \left(\frac{x_{\rm \fexxv}}{200} \right)^{1/2}  \ ,
%
%\dot{M}_{\rm out}(\textmd{\fexxv})  &\equiv& 4 \pi b m_p \int_r^{200}
%n(r,\theta) v_z(r,\theta) r dr \ ,
%\dot{M}_{\rm out}^{\rm (global)}  &\equiv& m_p \int_{z=z_c} d^2 x ~ n(r,\theta) v_z
%\sim \frac{4 \sqrt{2} G^2}{c^5} F B_o^2 M^2 r_{\rm out}^{1/2} \nonumber \\
%&\sim& 1.04 \Msun \left(\frac{F}{0.1}\right) \left(\frac{B_o}{10^4\textrm{G}}\right)^2
%\left(\frac{M}{4 \times 10^7\Msun}\right)^2 \left(\frac{r_{\rm out}}{10^6 R_o}\right)^{1/2}
%~ \textrm{year}^{-1}
\label{eq:Mdot}
\end{eqnarray}
where $x \equiv r/R_S$ and the upper limit of integration is indicative of the distance to
\fexxv\ location; i.e. $r_c = r_{\rm \fexxv}$ (see tab.~2).
%
%a value much higher than that needed to power its bolometric
%luminosity of $\sim 10^{45}$ erg/s.
%
%where $(r,\theta)$ here is the cylindrical coordinates and density
%and velocity profiles can be respectively split into radial $g(r)$
%and angular $f(\theta)$ components such as
%\begin{eqnarray}
%n(r,\theta) &=& n_o f_n(\theta) g_n(r) =  n_o \frac{R_S}{r} f_n(\theta) \ ,
%\\
%v_z(r,\theta) &=& v_{z,o} f_v(\theta) g_v(r) = 10^{-2}
%\left(\frac{R_S}{2 R_{\rm ISCO}}\right)^{1/2}
%\left(\frac{R_S}{r}\right)^{1/2} c  f_v(\theta)  \ ,
%%~ \textrm{year}^{-1}
%\label{eq:Mdot}
%\end{eqnarray}
%with $R_S = 3 \times 10^{13} M_8$ cm and the radius of the ISCO
%(innermost stable circular orbit) of $R_{\rm ISCO}=3 R_S$ for a
%non-spinning black hole. This leads to
%\begin{eqnarray}
%\dot{M}_{\rm out}^{\rm (global)}  &\sim& 6.58 \times 10^{25}  \pi m_p c n_o M_8^2
%\left(\frac{r_{\rm out}}{10^6 R_S} \right)^{1/2} f_n(\theta)
%f_v(\theta)  \ , \\
%&\sim&  2.16 \Msun ~ M_8^2 \left(\frac{r_{\rm out}}{10^6 R_S}\right)^{1/2} ~ \textmd{year$^{-1}$}\ ,
%\label{eq:Mdot}
%\end{eqnarray}
This value is consistent with the earlier estimate of $\sim 3
\Msun$~yr$^{-1}$ \citep[][]{Pounds03,PoundsPage06}. %{\bf Note that
%we assumed $r_{\rm out} \sim 10^6 R_S$ where the self-similar
%prescription could break down due to the interaction with the
%galaxy's spheroids at pc-scale. }
%
Since the corresponding local mechanical power is given by
\begin{eqnarray}
\dot{E}_{\rm out}^{\rm (local)}  \equiv \dot{M}^{\rm (local)}
v_{\rm out}^2 \propto r^{-1/2} \ ,
\end{eqnarray}
the local kinetic power of the \fexxv\ outflow is dominated by the
inner outflow radius $R_t$ yielding
\begin{eqnarray}
\dot{E}_{\rm out}(\textmd{\fexxv})  \sim \frac{1}{2} 
\dot M_{\rm out}(\textmd{\fexxv}) v_{\rm \fexxv}^2 \sim 2 \times 10^{44} ~ \textmd{erg~s$^{-1}$} \ ,
\end{eqnarray}
%
%assuming $r_{\rm in}/R_o \sim R_c({\rm \fexxv})/R_o \sim 100$ based
%on the current results in Table~2. This is indeed
a value comparable
to power the observed  X-ray luminosity $\sim 10^{44}$ erg~s$^{-1}$
\citep[][]{Pounds03}  also potentially providing a large impact on
AGN feedback process at large scales \citep[e.g.][]{CK12}. A
similarly large outflow power has been made, for example, to other
bright AGNs such as PDS~456 \citep[][]{Reeves03,Nardini15}.

\section{Summary \& Discussion}

We have demonstrated, by modeling its {\it XMM-Newton} spectrum,
that MHD-driven winds with $n \propto r^{-\alpha}, \, \alpha \simeq
1$, originally proposed to account for the X-ray WAs in Seyferts,
can also encompass the UFOs, {\it i.e. the high-velocity X-ray
absorbers} of {the} bright Seyfert \pg. The absorber's properties of
\pg\ as manifest by the \fexxv/\fexxvi\ transition properties, are
determined mainly by the wind mass flux $\dot{m}_{\rm a,o}$, the
disk temperature $kT_{\rm bbb}$ and the observer's viewing angle
$\theta$. By producing a grid of model K-shell Fe absorption lines,
appropriate to photoionized MHD winds, we found the that the
absorbers' physical conditions are well constrained by our models.
Thus the \fexxv\ and \fexxvi\ properties are respectively given by
the location of maximum opacity at $r_c/R_S = 234$ and $66$, LoS
velocity $v_c/c = 0.115$ and $0.208$, ionization parameter $\log
\xi_c = 5.31$ and $5.80$, { total H-equivalent columns } ${\cal
N}_H=1.21 \times 10^{23}$ cm$^{-2}$ and $1.67 \times 10^{23}$
cm$^{-2}$, with the wind truncated at radius $f_t \equiv R_t/R_S =
10^{1.48} \approx 30$. While the best-fit values of these parameters
are roughly consistent with the earlier analysis
\citep[e.g.][]{Gofford13}, our model can further provide a
geometrical and physical identification of the UFO in \pg\ data
rather than phenomenological interpretation as illustrated in
Figure~\ref{fig:fig6}b.
In fact, a recent discovery of the unambiguous P-Cygni-like profile of the UFOs made by  simultaneous X-ray observations with {\it XMM-Newton} and {\it NuSTAR} of a similar luminous quasar, PDS~456, indicates a similar spherically-extended wind geometry rather than a narrow collimated radial streamline \citep[][]{Nardini15} in consistence with our MHD-driven view discussed for \pg.

Although in this paper we are focusing on the origin of the detected
Fe K$\alpha$ UFOs, our model winds extend over a large range in $r,
\xi$ and $v$. As such, they imply the presence of other charge
states that contribute to the  Fe K$\alpha$ transition by including
\fexviii-\fexxiv. We found that should these additional states be
included in our analysis, the  Fe K$\alpha$ feature would have been
much broader than seen in the data. Given our fits of
Fig.~\ref{fig:fig4}a and Table~\ref{tab:tab2}, one must surmise that
the effective contribution to 1s-2p transition from \fexviii\
through \fexxiv\ in the data ought to be very small (if not none).
There are a number of remedies: (i) It is conceivable that the
intrinsically broad absorption feature due to ionized iron at all
charge states could be externally filled by scattered resonant line
photons which would suppress its otherwise broader
signature. Any continuous wind model will inevitably come across
this issue of the contribution of states other than highly-ionized
(e.g. H/He-like) ones. (ii) The radial wind density profile
might be steep enough to suppress the ionic column at large
distances. On the other hand, this solution may not be consistent
with the observed slow absorbers (i.e. warm absorbers) since they
originate from large distances in this model. (iii) It is also
probably that the fast absorbers (i.e. UFOs) could be a collection
of discrete (small) gas clouds along the LoS instead of a
large-scale continuous flow \citep[e.g.][]{Misawa14} that might also
be in a constant pressure equilibrium causing the suggested thermal
instability \citep[e.g.][]{HBK07}. While we note this long-standing
question, this is beyond the scope of our current study.

To compute the spectra of a truncated wind, we simply
removed the contribution to the line feature by the self-similar
section of the wind that originates at $R < f_t R_S$. We also
repeated the photoionization of a wind in which this section has
already been removed, assuming always that the ionizing source is
located at $r=0$. This second calculation produced similar results
with slightly smaller values for ${\cal N}_H$, because of the
slightly larger flux of ionizing radiation at the values of $\theta$
considered. This could be ameliorated by a slight increase in the
value of $n_o$.
The constrained truncation radius $R_t \approx 30 R_S$ in
Table~\ref{tab:tab2} is statistically favored in the context of our
MHD-wind model particularly so as to suppress the blue tail of the
absorption feature. On the other hand, \citet{Giustini12}, for
example, have considered a thermally-driven wind based on the model of
\citet{Luketic10}. They found a relatively sharp blue edge of the
line profile without truncation due to non-monotonic profiles for
wind streamlines and opacity along a LoS. This implies that a
complex geometry of the wind also needs to be further explored by
extending the model beyond self-similar limit. 

It is suggested from a long {\it Suzaku} observation that a similar
fast X-ray absorber (by iron K-shell transition at an implied
outflow velocity of $v \sim 0.25 c$) in PDS~456 exhibits rapid
variability as short as $\sim 1$ week \citep[][]{Gofford14}. Their
estimate of the absorber's location in PDS~456 ($r/R_S \sim 100 -
1800$)  is very similar to our estimate $r_c$ in \pg.
The current steady-state model is not appropriate for treating such
a time-variability in its absorption features, but it is conceivable
that the observed variable nature may be associated with the change
in wind density (perhaps resulted from the change in mass-loading)
and/or changing streamline configurations due to the variable
magnetic fields.
The model is in a good agreement with the data while there appears
to be additional weak (intrinsic) absorbers at higher energies
($\sim 8-9$ keV).  These weak absorption structures can be due to
the resonance series converging to the \fexxv\ edge
\citep[e.g.][]{Kallman04, T11a}. There can be also some
contamination due to the presence of the background (instrumental)
emission lines such as Cu K$\alpha$ at 8 keV in the EPIC-pn
spectrum.
%These are  unlikely to be another wind absorption component.

%Despite a simplistic treatment of MHD-driven outflows, our photoionization calculations show that the %characteristic wind variables listed in Table~2 can be  well explained in this toy model.
%
In this paper we employ a well-studied semi-analytic wind model as a
primary component. We feel that despite their simplicity, such
models should not be dismissed off hand compared to large scale
purely numerical simulations for a number of reasons: First, even
the state-of-the-art simulations  today have not yet provided {the}
practical and direct observables  addressed  in this paper at an
observationally-relevant level. Second, it is still computationally
extremely challenging to include self-consistently multi-scale,
multi-dimensional radiative transfer for plasma/atomic physics
necessary to simulate the kind of transitions seen in UV/X-ray data
while simultaneously covering a large spatial scale (say, ranging
from 10 Schwarzschild radii all the way out to pc-scale) without
suffering from numerical instability and boundary conditions
susceptibility.
%
%Our models are also not generalized enough to adequately address a
%bigger picture like an AGN feedback. It is thus beyond the scope of
%our current work although its importance is critical in AGN study.
%
In the next step a more self-consistent disk-wind morphology needs
to be considered by constructing a sophisticated (perhaps dynamical)
model \citep[e.g.][]{Ohsuga09,Ohsuga11} also incorporating  detailed
radiative transfer for spectral lines
\citep[e.g.][]{Kallman04,Garcia13}.

In this preliminary calculation we assumed $\dot m_{\rm a,o}=1$
corresponding to the density of matter on the disk surface $n_o = 5
\times 10^{11}$ cm$^{-3}$. We have used slightly different values
for $n_o$ and noted their weak influence of the end results. We note
that our assumed value is slightly higher than a { fiducial AGN}
value of $10^{10}$ cm$^{-3}$  \citep[e.g.][and references
therein]{CKG03, T11a, Gofford13} but its possible range can be
considered as broad as $10^{10} \lesssim n_o \lesssim 10^{17}$
cm$^{-3}$
%\textbf{{\em these high values are for disks of  galactic sources}}
\citep[e.g.][]{LaorNetzer89,GeorgeFabian91,Garcia13} and an accurate assessment
requires a more realistic modeling of accretion disk physics and its
response to photoionization process.

The present analysis is based on a selected fiducial wind structure
that we have examined in our earlier work (F14). Within the
three-parameter model spanned by $\theta, kT_{\rm bbb}$ and $R_t$ in
this paper, we do not notice degeneracy in the best-fit spectrum.
Considering a complexity of  magnetized disk-wind physics, however,
it is conceivable that we may find another best-fit solution from
slightly different wind conditions. Removing such potential
degeneracy is in principle challenging since there is little {\em a
priori} knowledge (at least observationally) of the underlying wind
structure. Nonetheless, it will be possible to rule out some of the
degenerate wind solutions by further including multiple ions of
different charge states both at soft X-ray transitions (below 3-4
keV) and Fe-K$\alpha$ transitions simultaneously since all these
absorption signatures should be coupled in the context of our
continuous disk-wind scenario. We thus plan to extend the current
preliminary model to include the soft X-ray WAs to examine a
coherent predictability of the model using those AGNs exhibiting
both WAs and UFOs.

We anticipate the upcoming missions such as {\it Astro-H} and {\it Athena} to contribute
significantly to this goal by providing more detail on the Fe-K
component of the wind as well as soft X-ray absorbers, and thus to further clarify our picture of
AGN structure.

\acknowledgments
We are grateful to the anonymous referee for the constructive criticism to improve the quality of the manuscript.

\end{document}